\documentclass[reprint,superscriptaddress,amsmath,amssymb,aps,nofootinbib,prx]{revtex4-1}

\usepackage{amsthm}
\usepackage{caption}
\usepackage{subfigure}
\usepackage[colorlinks=true,urlcolor=blue,citecolor=blue,linkcolor=blue]{hyperref}
\usepackage[shortlabels]{enumitem}
\usepackage{amsfonts}
\usepackage{graphicx}
\usepackage[caption = false]{subfig}
\usepackage{multirow}
\usepackage{float}
\usepackage{bbold}
\usepackage{stmaryrd}
\usepackage{color}
\usepackage{ragged2e}
\usepackage{hyperref}
\usepackage{verbatim}
\usepackage{cases}
\usepackage{amsfonts}
\usepackage{amssymb}
\usepackage[table]{xcolor}
\usepackage{epsfig}
\usepackage{bm}
\usepackage{setspace}
\usepackage{enumerate}
\usepackage{dcolumn}
\usepackage{bm}
\usepackage{enumitem}
\usepackage{caption} 
\usepackage{tikz}
\usepackage{multirow, array} 

\usepackage{graphicx}
\usepackage{dcolumn}
\usepackage{bm}
\usepackage{color}
\usepackage{amssymb}
\usepackage{lipsum}
\usepackage{amsmath}
\usepackage{graphicx}
\usepackage{array}
\usepackage{booktabs}
\usepackage[english]{babel}
\newcommand{\R}{\mathbb R}
\newcommand{\Q}{\mathbb Q}
\newcommand{\I}{\mathbb{R} \setminus \mathbb{Q}}
\newcommand{\Z}{\mathbb Z}

\newtheorem{theorem}{Theorem} 

\newtheorem{corollary}[theorem]{Corollary}

\makeatletter
\providecommand{\sf@counterlist}{}
\makeatother

\begin{document}

\title[Two-center problem with harmonic-like interactions]
{Two-center problem with harmonic-like interactions: periodic orbits and non-integrability}

\author{A.M. Escobar Ruiz}%
\email{admau@xanum.uam.mx}
\affiliation{Departamento de Física, Universidad Autónoma Metropolitana--Iztapalapa, P.O. Box  55--534, México, D.F., 09340 M\'exico}

\author{Lidia Jiménez--Lara}%
\email{lidia@xanum.uam.mx}
\affiliation{Departamento de Física, Universidad Autónoma Metropolitana--Iztapalapa, P.O. Box  55--534, México, D.F., 09340 M\'exico}

\author{J. Llibre}%
\email{jaume.llibre@uab.cat}
\affiliation{Departament de Matem\`{a}tiques, Universitat Aut\`{o}noma de Barcelona, 08193 Bellaterra, Barcelona, Catalonia, Spain}

\author{Marco A. Zurita}%
\affiliation{Departamento de Física, Universidad Autónoma Metropolitana--Iztapalapa, P.O. Box  55--534, México, D.F., 09340 M\'exico}


\begin{abstract}
We study the classical planar two-center problem of a particle $m$ subjected to harmonic-like interactions with two fixed centers. For convenient values of the dimensionless parameter of this problem we use the averaging theory for showing analytically the existence of periodic orbits bifurcating from two of the three equilibrium points of the Hamiltonian system modeling this problem. Moreover, it is shown that the system is  generically non-integrable in the sense of Liouville–Arnold. The analytical results are complemented by numerical computations of the Poincaré sections and Lyapunov exponents. Explicit periodic orbits bifurcating from the equilibrium points are presented as well.
\end{abstract}

\maketitle

\section{Introduction}

Nonlinear dynamical systems are central objects in theoretical physics. For instance, in classical mechanics \cite{Landau, GPS2002, Arnold89, AKN2006} appears the rich and complex behaviour of physical relevant systems such as the 3-body problem, dynamical astronomy, Henon-Heiles Hamiltonian systems, rigid bodies problems, and non-linear Hamiltonian systems \cite{Conto, Henon, MacKay}.

Among the various trajectories that mechanical systems can exhibit, periodic solutions play a fundamental role. These trajectories represent a closed motion in the phase space, offering valuable insights into the underlying dynamical properties and stability of the system. Needless to say that, in general, the intricate time evolution of non-linear systems do not admit a straightforward analysis. There is a lack of universally applicable formulas for determining periodic trajectories in dynamical systems. Therefore, the development of asymptotic approximations to the solutions of a nonlinear differential system as well as their numerical approaches are needed.

In this context the averaging theory formulated in Fatou's seminal work \cite{Fatou} offers a systematic approach to extract essential information from complex dynamical systems. Subsequent contributions in the 1930s by Bogoliubov and Krylov \cite{Bogoliubov1934}, as referenced by Bogoliubov \cite{Bogoliubov1945} in 1945, significantly increased both practical applications and theoretical understanding of the averaging theory.

Over time, the ideas of averaging theory have undergone refinement and expansion in various directions, catering to both finite and infinite-dimensional differentiable systems. For contemporary literature and developments in averaging theory, we refer the readers to the works of Sanders, Verhulst, Murdock (see \cite{Sanders} and references therein), and Verhulst \cite{Verhulst}, among others, which provide modern expositions and present-day results on the subject.

The fundamental premise of the averaging theory lies in the recognition that many physical systems exhibit fast and slow motions simultaneously. By exploiting this timescale separation, averaging techniques aim to construct simplified models that capture the essential dynamics while filtering out fast oscillations and transient behavior. This reduction in complexity not only facilitates analytical tractability but also provides qualitative insights into the long-term behavior of the system. Concrete applications can be found in the works \cite{Adriana,Llibre_2011}.

In this paper we aim to investigate the dynamics of a two-center problem with harmonic interactions using the averaging theory. This system can be viewed as a limiting case of the 3-body harmonic oscillator, a nine-parameter system with three arbitrary masses, three rest lengths, and three spring constants.

In the simplest scenario involving equal masses on the plane and equal spring constants, the 3-body harmonic oscillator exhibits a remarkably diverse dynamics in function of the energy \cite{Katz2019, 3bodyexperimental}. This diversity manifests in a power-law statistics reminiscent of the Levi-walk model \cite{Katz2019}. Even when restricted to the invariant manifold of zero total angular momentum, the parameter space displays regions of both regular and chaotic dynamics due to inherent non-linearities stemming from non-zero rest lengths \cite{Katz2020}. Upon setting these rest lengths equal to zero, the system attains superintegrability.

It is noteworthy that at zero rest lengths, the corresponding classical and quantum 3-body oscillator system becomes exactly solvable \cite{Turbiner2020}. However, for non-zero rest lengths, an exact solution is not known. Consequently the quantum 3-body harmonic oscillator \cite{Olivares-Pilon_2023} serves as a practical model for testing theoretical and numerical methodologies aimed at elucidating the interplay between classical and quantum mechanics within chaotic systems.

In the case when two bodies are considered infinitely massive, the 3-body harmonic oscillator reduces to the two-center problem the one investigated in this paper. Despite this simplification, we will see that the two-fixed-centers problem inherits the chaotic behavior of the original system. Numerical as well as analytical tools based on the averaging theory are used to explore the dynamics of this system. The main goal is to find periodic trajectories emanating from the equilibrium points of this system.
\vspace{-10pt}
\subsection{The two-center problem with harmonic-like interactions}

In the Euclidean space $\R^2$ we consider a two-center problem of a non-relativistic point particle $m$ subjected to harmonic-like interactions with two fixed centers possessing the same {\it constant of elasticity} $k>0$. In cartesian coordinates $(X,Y)$ the Hamiltonian of the system is of the form:
\begin{equation}\label{H}
{\mathcal H }\ = \ \dfrac{1}{2\,m} \left(\,P_X^2 \ + \ P_Y^2\,\right)\ + \ V(X,Y) \ ,
\end{equation}
where $V(X,Y)$ is the translational-invariant potential
\begin{equation*}
\begin{split}
V(X,Y) \  &= \ \dfrac{1}{2} \,k\,  \big(\, \left(R_{1}-A \right)^2 \ + \ \left(R_{2}-A \right)^2 \,\big) \ , \\
&= \ \dfrac{1}{2}\, k \left(\sqrt{(X+L )^2+Y^2}-A\right)^2
\\ & \ + \ \left(\sqrt{(X-L)^2+Y^2}-A\right)^2 \,\big) \ ,
\end{split}
\end{equation*}
$R_1$, $R_2$ are the distances from the mass $m$ to the two fixed centers, respectively, which we assume are located at $(\pm L,0)$, and the constant $A\ge 0$ denotes the equilibrium distance from $m$ to each one of the fixed centers. Hence, the phase space is four-dimensional.  First, for the non-dimensionalization of $ {\mathcal H }$, we divide the expression (\ref{H}) by $m\,L^2\omega^2$, where $\omega^2 = k/m$, and define a dimensionless time $\tau=t\, \omega$. More precisely, we introduce the set of non-dimensional quantities:
{\small
\begin{equation*}
\begin{split}
x &=X/L \ , \quad y=Y/L \ , \quad   a=A/L \, , \quad   r_i=R_i/L \ , \quad \tau = t \,\omega  \ , \\
p_x &= \dfrac{dx}{d\tau} \ , \quad p_y= \dfrac{dy}{d\tau} \ , \quad H ={\mathcal H}/(m\,L^2\omega^2) \ , \ U=V/(m\,L^2\omega^2) \ .
\end{split}
\end{equation*}
}
In these variables the original Hamiltonian (\ref{H}) is written in dimensionless form as follows
\begin{equation}\label{H1}
H\ = \ \dfrac{1}{2} \left(\,p_x^2 \ + \ p_y^2\,\right)\ + \  U(x,y) \ ,
\end{equation}
\begin{equation*}
\begin{split}
\label{U}
U(x,y) \  = \ \dfrac{1}{2}  \bigg(\, \left(\sqrt{(x+1)^2+y^2}\,-\,a \right)^2
\\
+\left(\sqrt{(x-1)^2+y^2}\,-\,a \right)^2 \, \bigg) \ ,
\end{split}
\end{equation*}
here the only remaining dimensionless parameter is $a$. Below in Fig. \ref{F1}, the geometrical settings of the system are presented in detail, and in Fig. \ref{figure2} we graph the potential.

In the special case $a=2$, a configuration of equilibrium $r_1=r_2=a$ corresponds to the equilateral triangle with sides $(2,2,2)$, where the particle and the two centers mark the vertices.

\begin{figure}[h]
	\includegraphics[width=8.0cm]{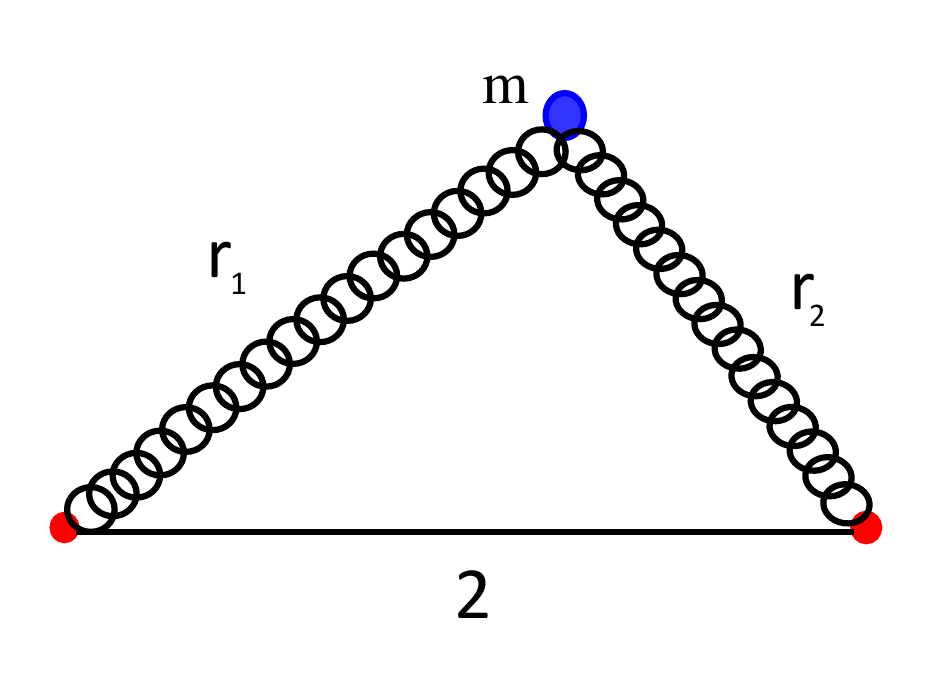}
	\caption{Planar two-center problem with harmonic-like interactions in dimensionless variables.  The distance between the two fixed centers is 2 and the only free parameter is $a=A/L$. The reference system with the origin in the midpoint of the line which connects the two centers is adopted. These centers are located at $(\pm 1,\,0)$, respectively.}
	\label{F1}
\end{figure}

\textbf{Remark}: At $a=0$ system (\ref{H1}) coincides with the 2D isotropic harmonic oscillator, a superintegrable system possessing three algebraically independent first integrals in the Liouville sense.

Along the line $x=0$, for $0 \leq a \leq 1$ the potential (\ref{H1}) possesses a critical point (a minimum) at $y=0$, $U(0,0)={(a-1)}^2$, whilst for $a>1$ this point $(0,0)$ becomes a maximum and two symmetric minima located at $y_{\pm}=\pm \sqrt{a^2-1}$, respectively, occur with $U(0,y_{\pm})=0$.

On the line $y=0$, for $0 \leq a \leq 1$ the potential (\ref{H1}) displays a minimum at $x=0$, whereas for $a>1$ two additional  symmetric maxima located at $x=\pm 1$, respectively, emerge. Also, for any value of $a$ the derivative of the potential is discontinuous at $x=\pm 1$.

\begin{figure*}
	\centering
	\subfigure[]{\includegraphics[width=0.31\textwidth]{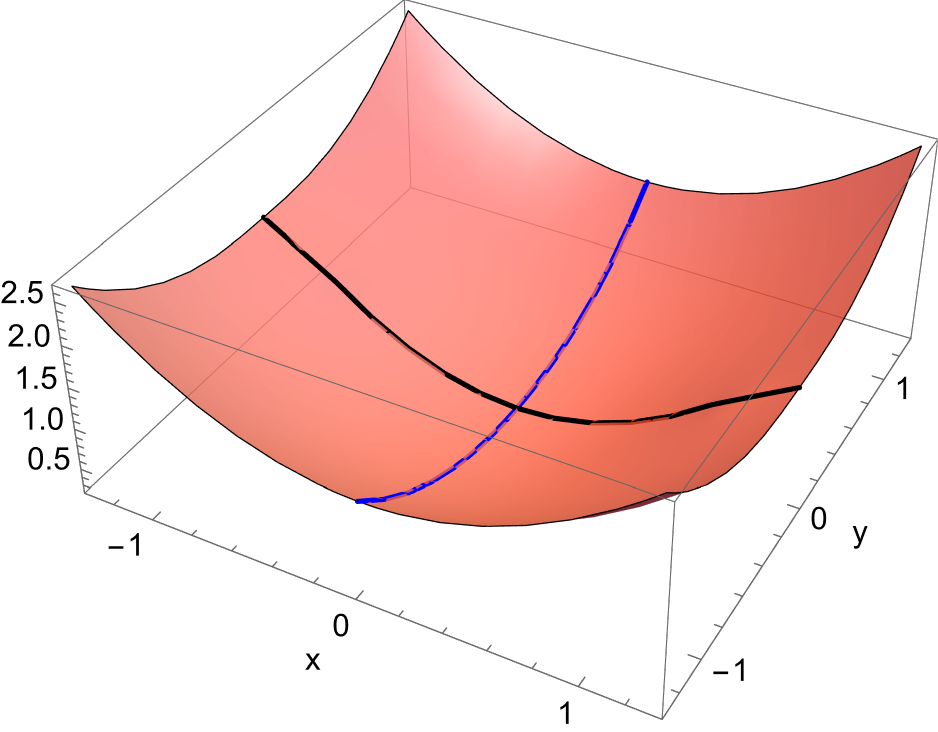}}
	\quad
	\subfigure[]{\includegraphics[width=0.31\textwidth]{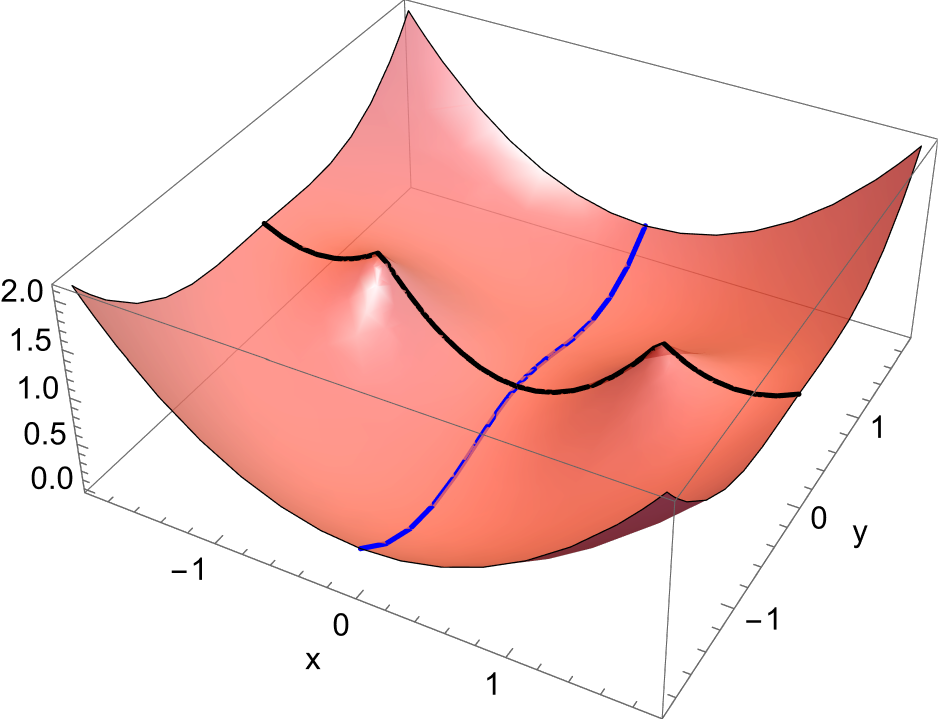}}
	\subfigure[]{\includegraphics[width=0.31\textwidth]{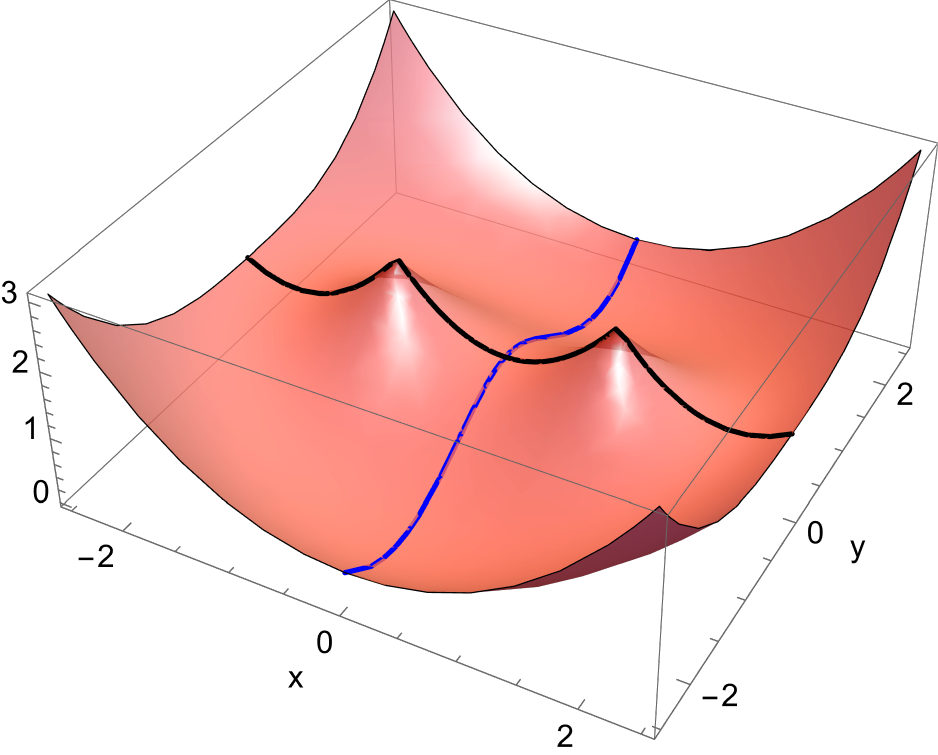}}\\
	\subfigure[]{\includegraphics[width=0.26\textwidth]{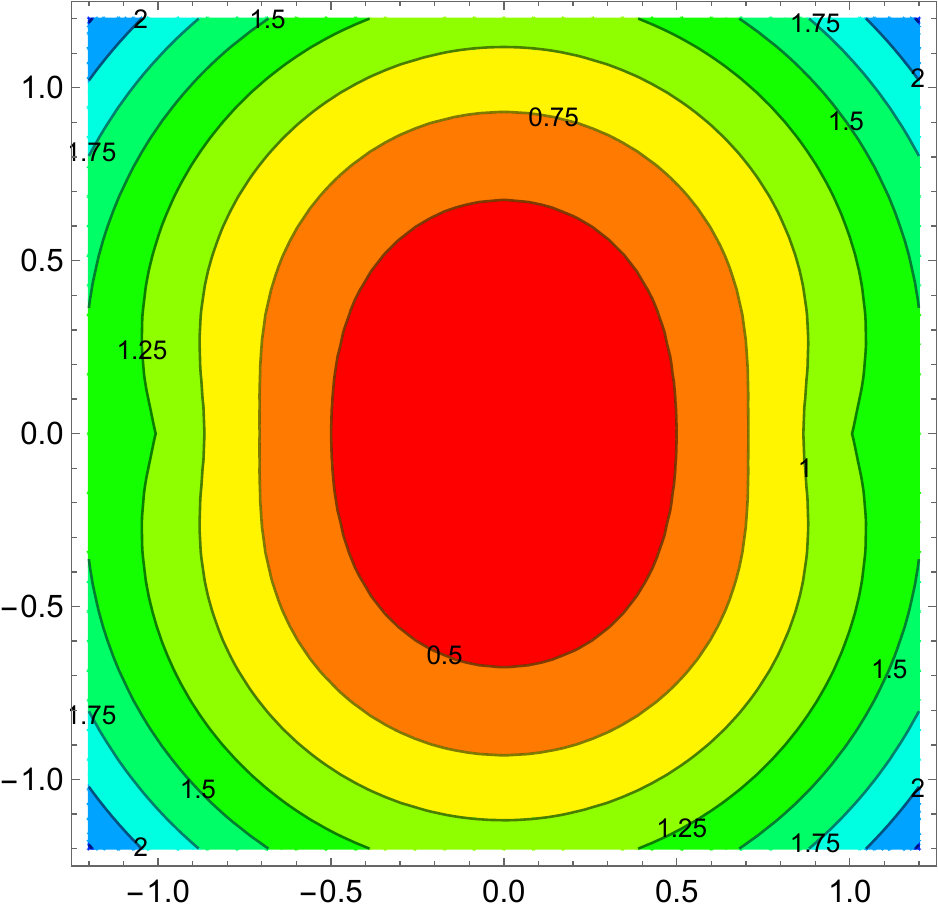}} \hfill
	\subfigure[]{\includegraphics[width=0.26\textwidth]{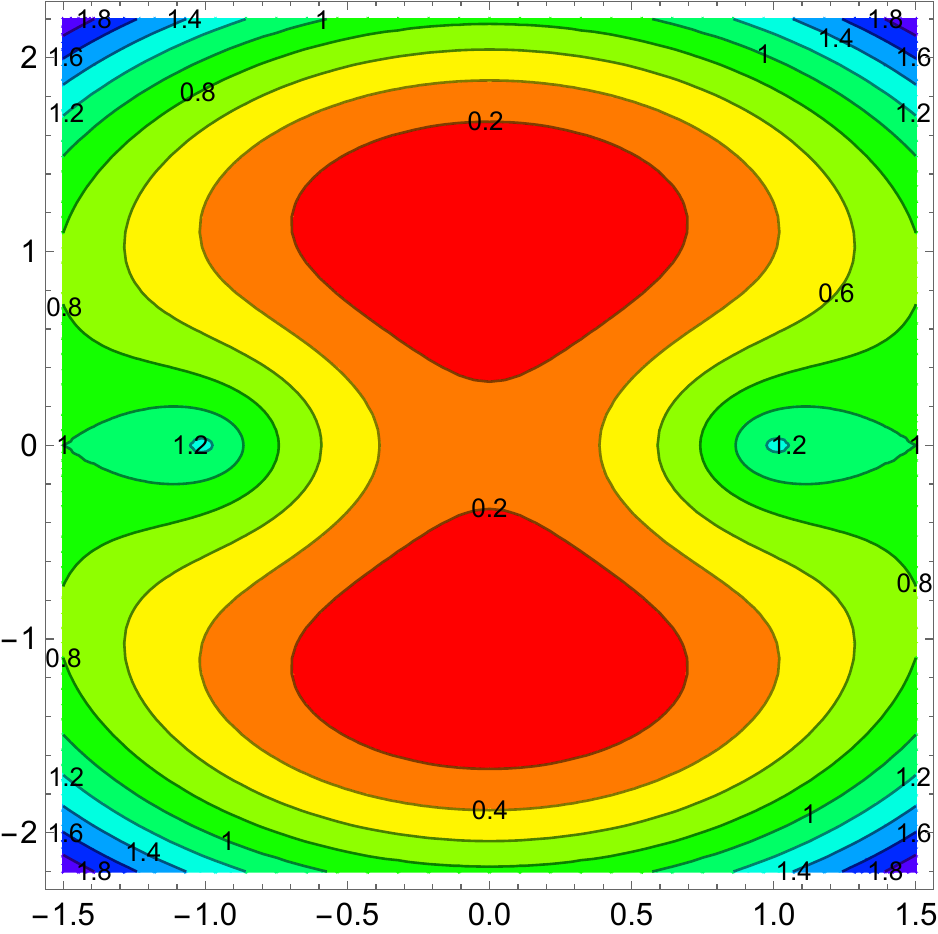}} \hfill
	\subfigure[]{\includegraphics[width=0.26
		\textwidth]{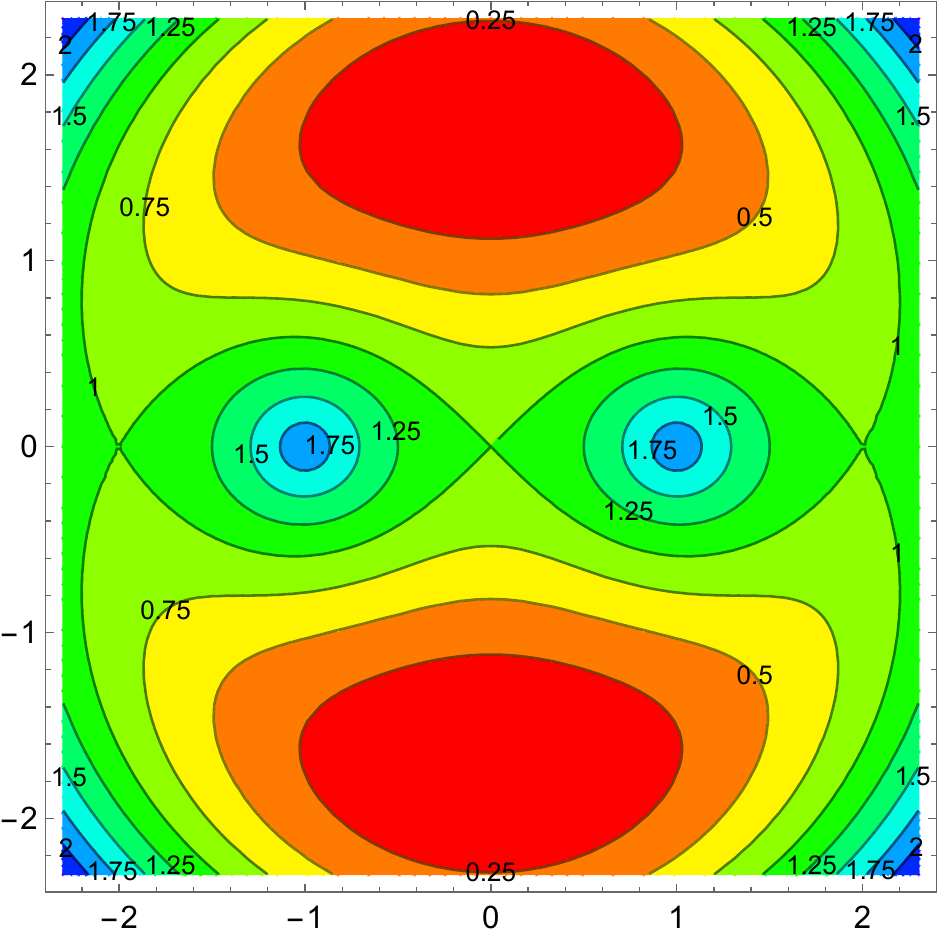}}
	\caption{The potential $U(x,y)$ in (\ref{H1}) as a function of the parameter $a$; (a) $a=\frac{1}{2}$, (b) $a=\frac{3}{2}$, (c) $a=2$. The corresponding level curves are shown in Figs. (d), (e) and (f), respectively.}
	\label{figure2}
\end{figure*}

For the Hamiltonian (\ref{H1}) the associated Hamilton's equations of motion are
\begin{equation}\label{dotx}
\begin{aligned}
    \dot{p}_x \  &=  - 2\,x+a \left( \dfrac{x+1}{\sqrt{(x+1)^2+y^2}} +  \dfrac{x-1}{\sqrt{(x-1)^2+y^2}} \right) , \\
    \dot{p}_y\  &=  - 2\,y+a\,y \left( \dfrac{1}{\sqrt{(x+1)^2+y^2}} +  \dfrac{1}{\sqrt{(x-1)^2+y^2}} \right) \ ,
    \\
    \dot{x}\ &= \ p_x \qquad , \qquad
    \dot{y}\ = \ p_y \ .
\end{aligned}
\end{equation}
This differential system is the main object of study of the present paper. It is invariant under the discrete symmetries
$$
\begin{array}{l}
\mathcal{S}_1: \ (x \rightarrow -x,\,y \rightarrow y , p_x \rightarrow -p_x,\,p_y \rightarrow p_y ),\\
\mathcal{S}_2: \  (x \rightarrow x,\,y \rightarrow -y , p_x \rightarrow p_x,\,p_y \rightarrow -p_y ),
\end{array}
$$
and $\mathcal{S}_1 \circ \mathcal{S}_2$. Then the orbits are symmetric with respect to the planes $(x,0,p_x,0)$ and $(0,y,0, p_y)$ as well as from the origin under the symmetry $\mathcal{S}_1 \circ \mathcal{S}_2$.

The paper is structured as follows. In section \ref{s2},  we show the Poincaré sections of the flow of the Hamiltonian system \eqref{dotx} and the maximum Lyapunov exponents for the values of the parameter $a=3/2,2,\sqrt{5},5$ and for some fixed values of the energy $H=E$. These Poincaré sections provide some information about the global dynamics of the Hamiltonian system \eqref{dotx}. 

In section \ref{s3} we recall the basic results of the averaging theory of first order for computing periodic orbits that we shall need for proving Theorem \ref{t1}. As we have said in section \ref{s4} we prove Theorem \ref{t1}.

In section \ref{s5} we compute periodic orbits of Theorem \ref{t1} for the values of the parameter $a=\sqrt{13}/3, \sqrt{5}, \sqrt{29}/2$. The case $a=\sqrt{5}$ which is excluded in Theorem \ref{t1}, is addressed in this section, and the periodic orbits are explicitly exhibited and graphed.

Finally, in section \ref{s6} we prove Theorem \ref{t2}.

\vspace{-0.5cm}

\subsection{Main results}

For $a >1$, the system (\ref{dotx}) has five equilibrium points ${\mathbf x}_0 \,\equiv\, (x,y,p_x,p_y)$, namely:
\[
(0,\,0,\,0,\,0), \quad (0,\,\pm \sqrt{a^2-1},\,0,\,0), \quad (\pm a,\,0,\,0,\,0).
\]
For $a>1$,  the eigenvalues of the equilibrium point $(0,\,\sqrt{a^2-1},\,0,\,0)$ are purely imaginary, i.e. they are $\pm i\, \sqrt{2}/a$ and $\pm i \sqrt{2(a^2-1)}/a$, and in the next theorem we describe the periodic orbit that bifurcate from this equilibrium.

\begin{theorem}\label{t1}
For $a>1$, $a\neq\sqrt{N^2+1}$ for all integer $N$ and $a\neq\sqrt{3}$,  in each energy level $H=h$ with $h>0$ sufficiently small, from the equilibrium point $(0,\,\sqrt{a^2-1},\,0,\,0)$ of the Hamiltonian system \eqref{dotx} can bifurcate one or more periodic orbits $(x(t,\varepsilon),y(t,\varepsilon), p_x(t,\varepsilon),p_y(t,\varepsilon))$ with initial conditions $(x(0,\varepsilon),y(0,\varepsilon), p_x(0,\varepsilon),p_y(0,\varepsilon))$ of the form
\begin{equation*}
\left(\varepsilon\tilde r,\, \sqrt{a^2-1}+\varepsilon\tilde\rho\cos \frac{\sqrt{2} g} {a}\tilde s ,\, 0,\, - \varepsilon\dfrac{\sqrt{2} g}{a} \tilde \rho \sin \frac{\sqrt{2} g} {a} \tilde s\right) + O(\varepsilon^2),
\end{equation*}
when the determinant of the Jacobian matrix 
\begin{equation}\label{jacobian}
\left. \frac{\partial (f_1(\rho,s),f_2(\rho,s))}{\partial(\rho,s)}\right|_{\rho=\tilde \rho,s=\tilde s}\ne 0\ .
\end{equation}
Here $\varepsilon>0$ is a small parameter and $g\equiv \sqrt{a^2-1}$. If $g\in \Q $, the orbit is periodic, and if $g\in \I$, the orbit is quasiperiodic. The values of the functions $f_i(\rho,s)$ for $i=1,2$, of the parameter $a$ and of the constants $\tilde r$, $\tilde \rho$, $\tilde s$ are given in the proof of this theorem.
\end{theorem}

Theorem \ref{t1} is proved in section \ref{s4}. In section \ref{s5} for different values of the parameter $a$ we prove the existence of one or two periodic orbits given by Theorem \ref{t1}.

Note that when $\varepsilon\to 0$ the periodic orbit of Theorem \ref{t1} bifurcates from the equilibrium point $(0,\,\sqrt{a^2-1},\,0,\,0)$.
Due to the $\mathcal{S}_2$ symmetry by studying the periodic orbit bifurcating from the equilibrium $(0,\,\sqrt{a^2-1},\,0,\,0)$ we are also studying the periodic orbit bifurcating from the symmetric equilibrium $(0,\,-\sqrt{a^2-1},\,0,\,0)$.   

If the periodic orbit obtained in Theorem \ref{t1} is not invariant under the $\mathcal{S}_1$ symmetry of the differential system \eqref{dotx}, there is another symmetric periodic orbit distinct to the one given in Theorem \ref{t1} with the initial condition $(-x(0,\varepsilon), \allowbreak y(0,\varepsilon), \allowbreak -p_x(0,\varepsilon), \allowbreak p_y(0,\varepsilon))$
 which is also near the equilibrium point $(0,\,\sqrt{a^2-1},\,0,\,0)$.

On the other hand, if 
$$C=\left(\tilde \rho\cos \frac{\sqrt{2} g} {a}\tilde s \right)^2+ \left(\dfrac{\sqrt{2} g}{a} \tilde \rho \sin \frac{\sqrt{2} g} {a} \tilde s\right)^2\ne 0\,,$$ then the $\mathcal{S}_2$ symmetry of the differential system \eqref{dotx} provides another periodic orbit distinct to the one given by Theorem \ref{t1} with initial conditions $(x(0,\varepsilon), \allowbreak -y(0,\varepsilon), \allowbreak p_x(0,\varepsilon), \allowbreak -p_y(0,\varepsilon))$, however this is not near the original equilibrium point  $(0,\sqrt{a^2-1},\,0,\,0)$, but near  $(0,-\sqrt{a^2-1},\,0,\,0)$. 

Finally, if $\tilde r C\ne 0$, then the $\mathcal{S}_1\circ \mathcal{S}_2$ symmetry of the differential system \eqref{dotx} provides another periodic orbit distinct to the one given in Theorem \ref{t1} with the initial condition $(-x(0,\varepsilon),-y(0,\varepsilon),-p_x(0,\varepsilon),-p_y(0,\varepsilon))$, and consequently near the equilibrium point $(0,-\sqrt{a^2-1},\,0,\,0)$. 

There is a local symmetry which inverts $y$ and $p_y$ around the equilibrium point  $(0,\sqrt{a^2-1},\,0,\,0)$, valid only when $\varepsilon\rightarrow 0$. In this way, for each periodic orbit bifurcating from the equilibrium point, there can be four periodic orbits bifurcating simultaneously from it. 

In short, we have proved the next corollary.

\begin{corollary}\label{c1}
Under the assumptions of Theorem $\ref{t1}$ and for each periodic orbit bifurcating from the equilibrium $(0,\,\sqrt{a^2-1},\,0,\,0)$, if $\tilde r C\ne 0$ and $a>1$, then at each energy level $H=h$ of the Hamiltonian system \eqref{dotx} with $h>0$ sufficiently small, there are at least $4$ periodic orbits, $2$ near the equilibrium $(0,\,\sqrt{a^2-1},\,0,\,0)$, and the other two near the equilibrium $(0,-\sqrt{a^2-1},\,0,\,0)$.
\end{corollary}

Of course, the integrable and non--integrable Hamiltonian systems can have infinitely many periodic orbits. In general, it is not easy to find explicitly a whole family of analytical periodic orbits mainly when the Hamiltonian system is non--integrable.  Here we find them in Theorem \ref{t1} and in Corollary \ref{c1}. Once we have proved that at any positive energy level sufficiently small there exist analytic periodic orbits, we can use them to prove the next result about the non--integrability of the Hamiltonian system \eqref{dotx} in the sense of Liouville--Arnold. A Hamiltonian system with $n$ degrees of freedom is called Liouville integrable if it possesses $n$ independent, mutually commuting integrals of motion that are in involution (i.e., their Poisson brackets are zero).

\begin{theorem}\label{t2}
Suppose that the Hamiltonian system \eqref{dotx} satisfies the hypotheses of Theorem $\ref{t1}$ and Corollary $\ref{c1}$. Then one of the following two statements hold: 
\begin{itemize}
\item[(a)] either the Hamiltonian system \eqref{dotx} is
Liouville--Arnold integrable and the gradients of the two constants
of motion are linearly dependent on some points of the periodic
orbits found in Theorem \ref{t1},
\item[(b)] or the Hamiltonian system \eqref{dotx} is not Liouville--Arnold
integrable with any second first integral of class $\mathcal{C}^1$.
\end{itemize}
\end{theorem} 

\vspace{-10pt}
\section{Poincaré sections and Lyapunov exponents}\label{s2}

Now for the Hamiltonian (\ref{H1}) we present the Poincaré sections on the $(y,p_y)$ plane, considering the values $a=3/2,2,\sqrt{5},5$ as a function of the energy $E=H$. The configuration of equilibrium $r_1=r_2=a$ with $a=2$ corresponds to the equilateral triangle with sides $(2,2,2)$. The value $a=\sqrt{5}$ was selected because this case was handled separately in the averaging theory whereas $a=5$ corresponds to a \textit{large} value of the parameter $a$. For $a>1$, the system has three equilibrium points (otherwise, it has only one). At $a=2$, the equilibrium configuration of the system corresponds to an equilateral triangle, which is why we have chosen the midpoint $a=\frac{3}{2}$. The Poincaré sections are determined from the intersection of trajectories, associated with given initial conditions within the phase space, with a lower-dimensional subspace ($x=0$, $y$, $p_x(y,p_y,E)$, $p_y$). They are transversal planes to the flow of the Hamiltonian system, and they can be regarded as a discretized version of the dynamical system retaining relevant properties of the original continuous system but acting in a reduced phase space. 
For the calculations of the Poincaré sections we take as a reference point of energy the value $E_s$ of the potential (\ref{H1}) evaluated at the saddle point $(0,0)$, namely $E_{s} \ = \ U(0,0) \ = \ {(a-1)}^2$. 

\begin{figure*}[h!]
\centering
\subfigure[$E=\frac{1}{2}E_s$]{\includegraphics[width=0.3\textwidth]{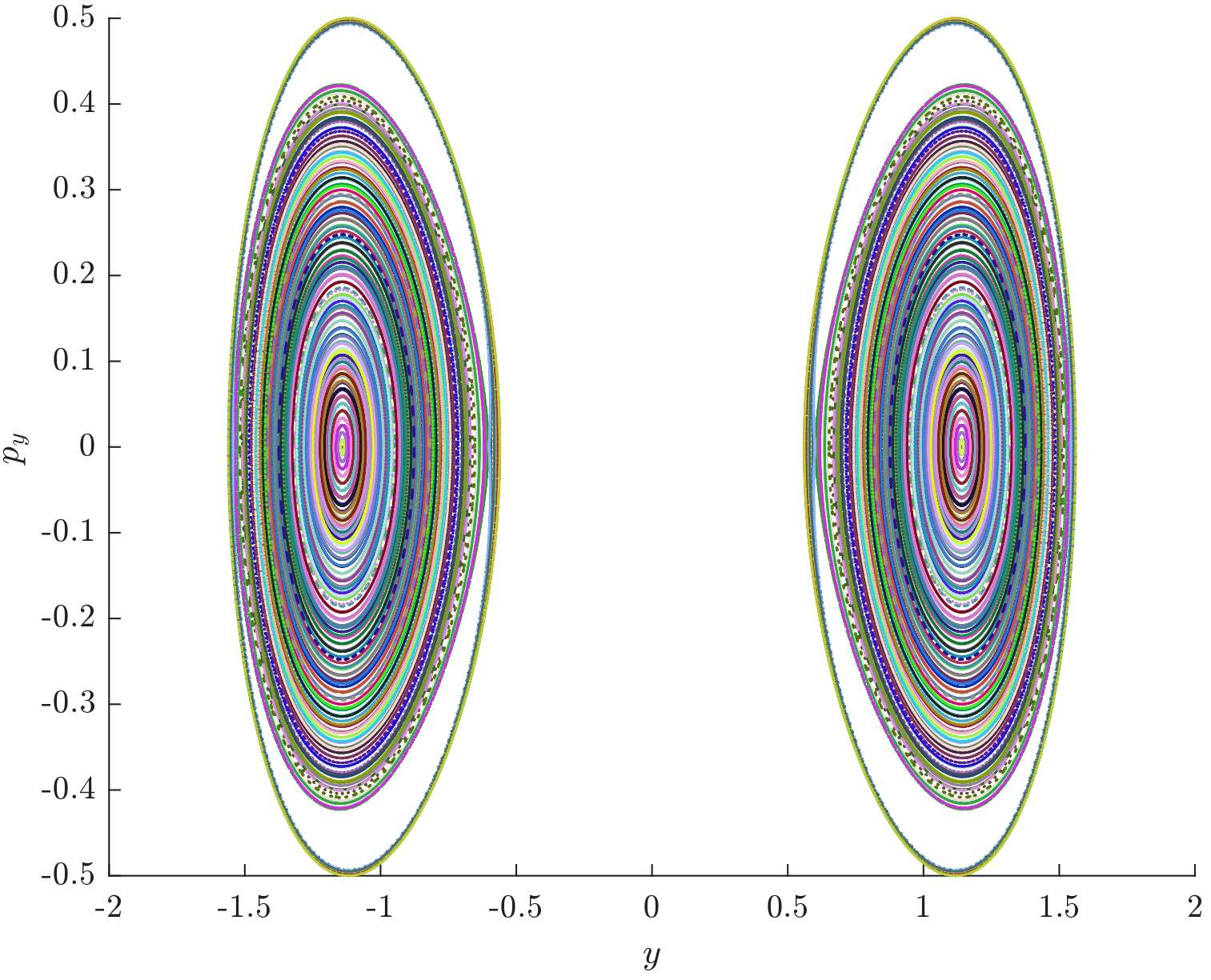}}
\hfill
\subfigure[$E=\frac{99}{100}E_s$]{\includegraphics[width=0.3\textwidth]{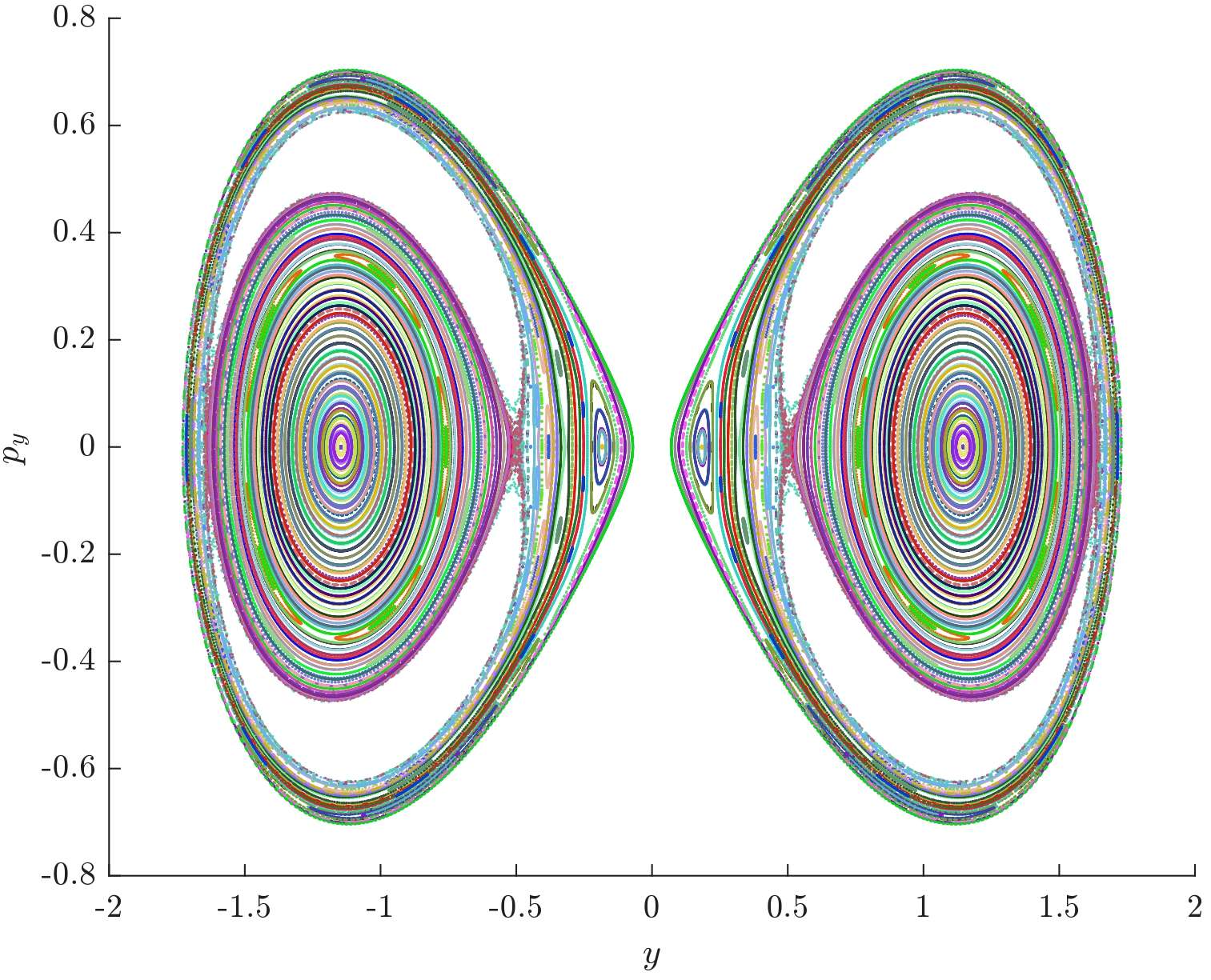}}
\hfill
\subfigure[$E=4\,E_s$]
{\includegraphics[width=0.3\textwidth]{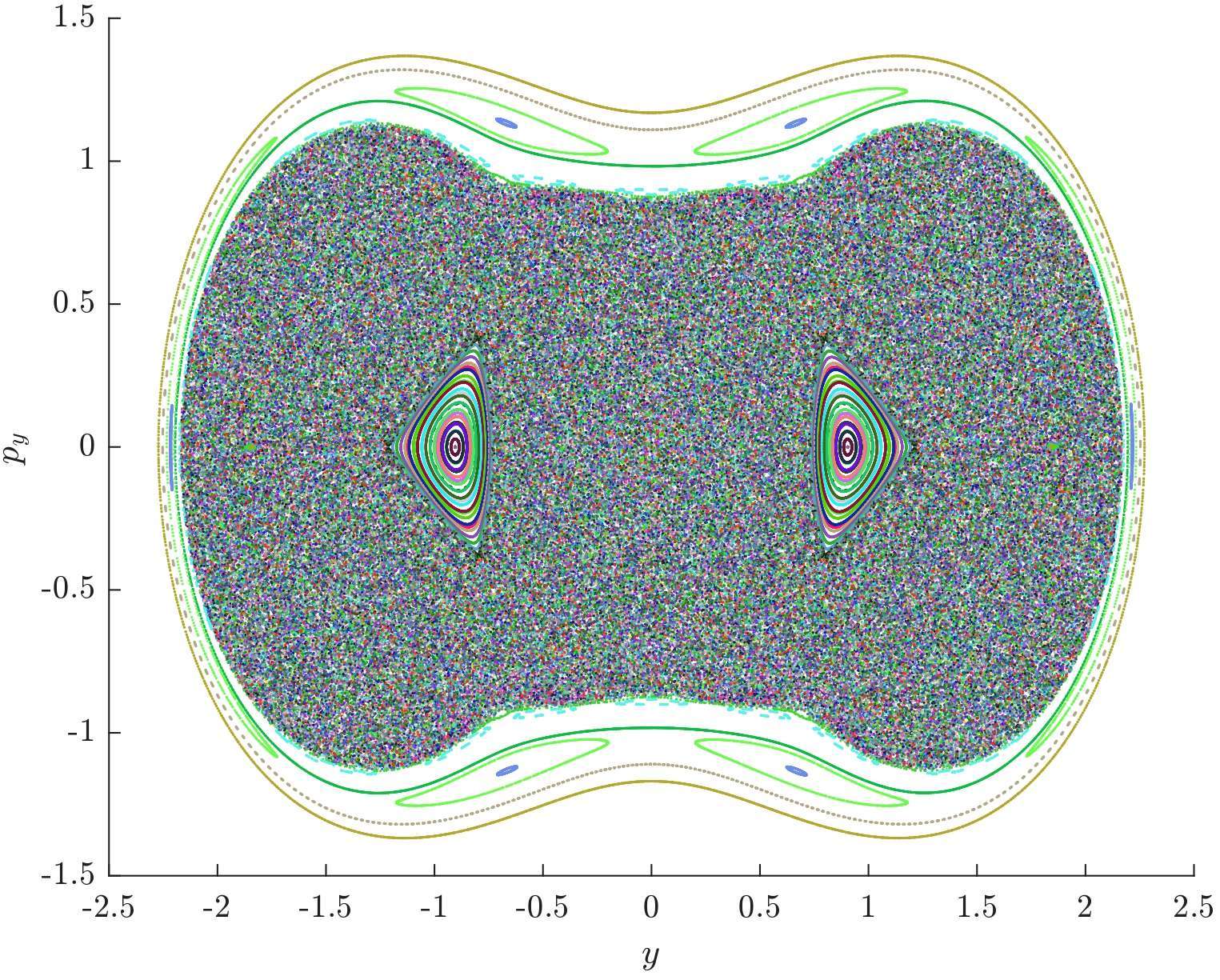}}
\\
\subfigure[$E=\frac{1}{2}E_s$]{\includegraphics[width=0.32\textwidth]{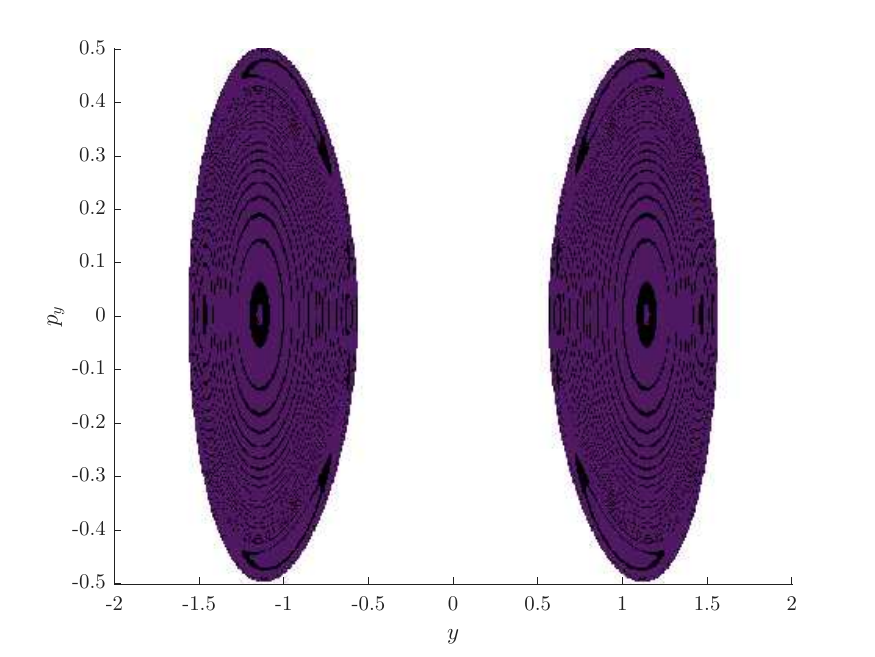}}
\hfill
\subfigure[$E=\frac{99}{100}E_s$]{\includegraphics[width=0.32\textwidth]{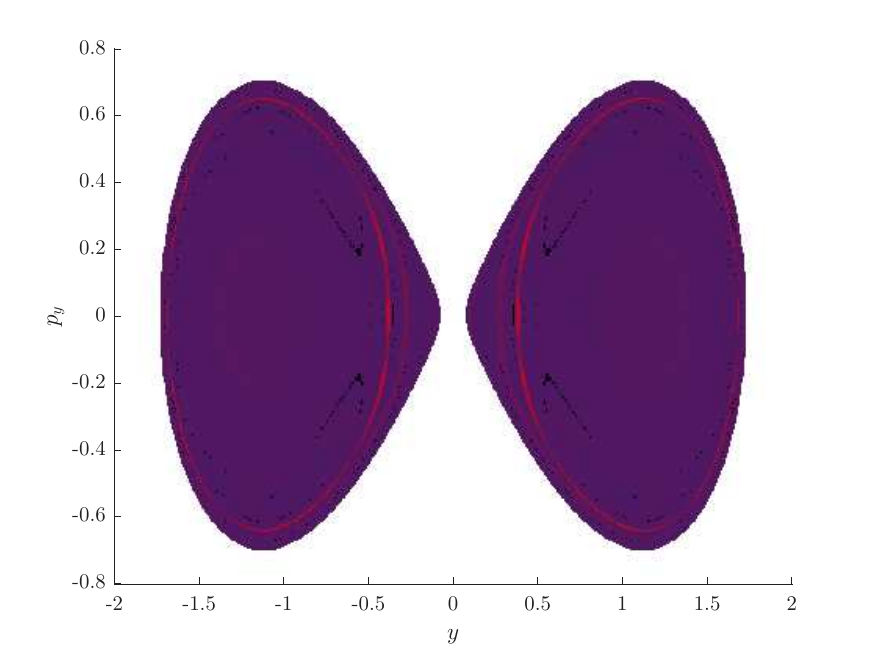}}
\hfill
\subfigure[$E=4\,E_s$]{\includegraphics[width=0.32\textwidth]{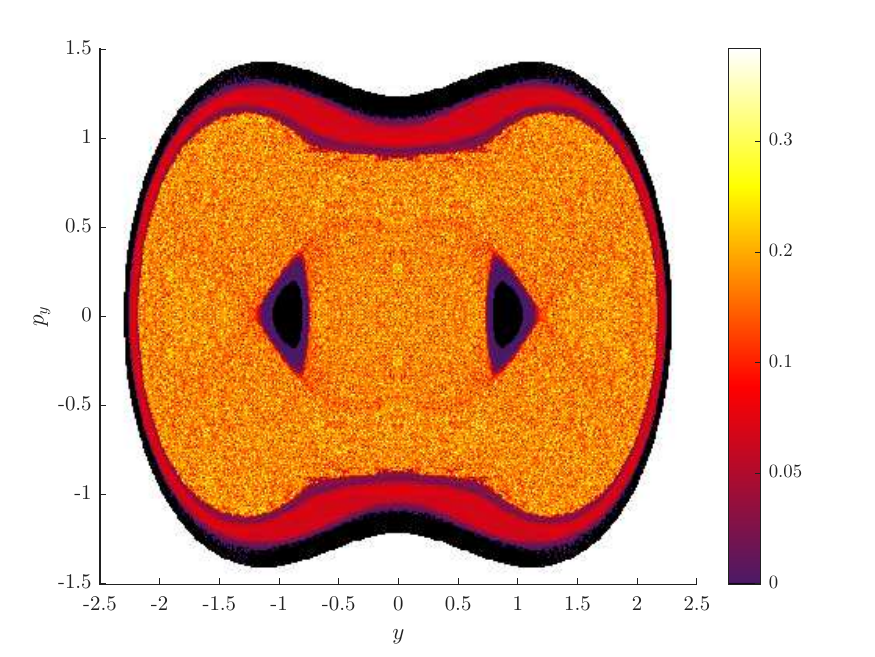}}
\caption{\small (a)-(c) oriented Poincaré sections, on the plane $(y,\,p_y)$, for $H$ (\ref{H1}) with $a=\frac{3}{2}$, at different values of the energy $H=E$; (d)-(f) the corresponding largest Lyapunov exponents. The equilibrium points are $(0,\,\pm \sqrt{5/4},\,0,\,0)$ and $E_s=1/4\,.$}
\label{Fa32}

\end{figure*}
\begin{figure*}[h!]
	\centering
	\subfigure[$E=\frac{1}{2}E_s$]{\includegraphics[width=0.3\textwidth]{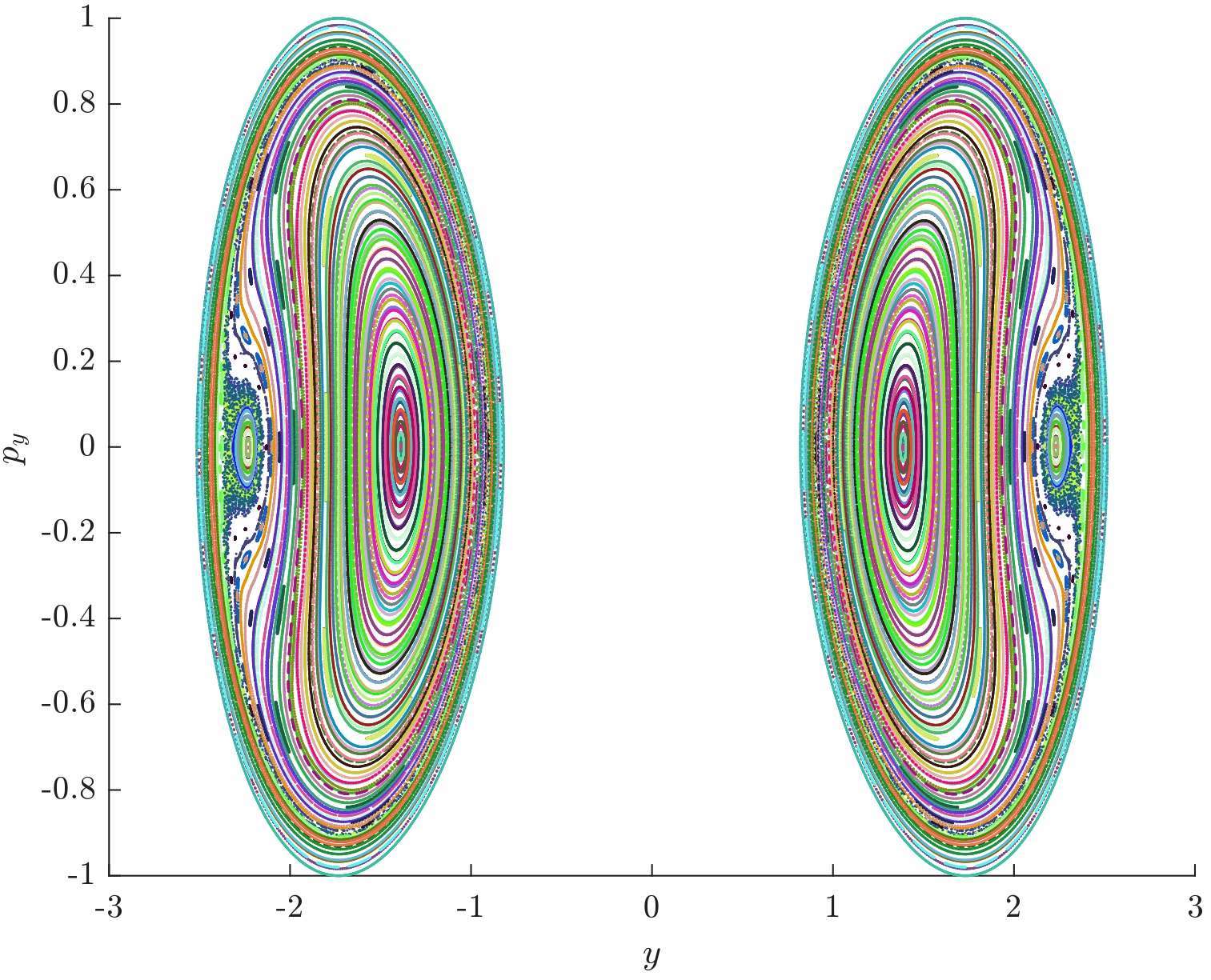}}
	\hfill
	\subfigure[$E=\frac{99}{100}E_s$]{\includegraphics[width=0.3\textwidth]{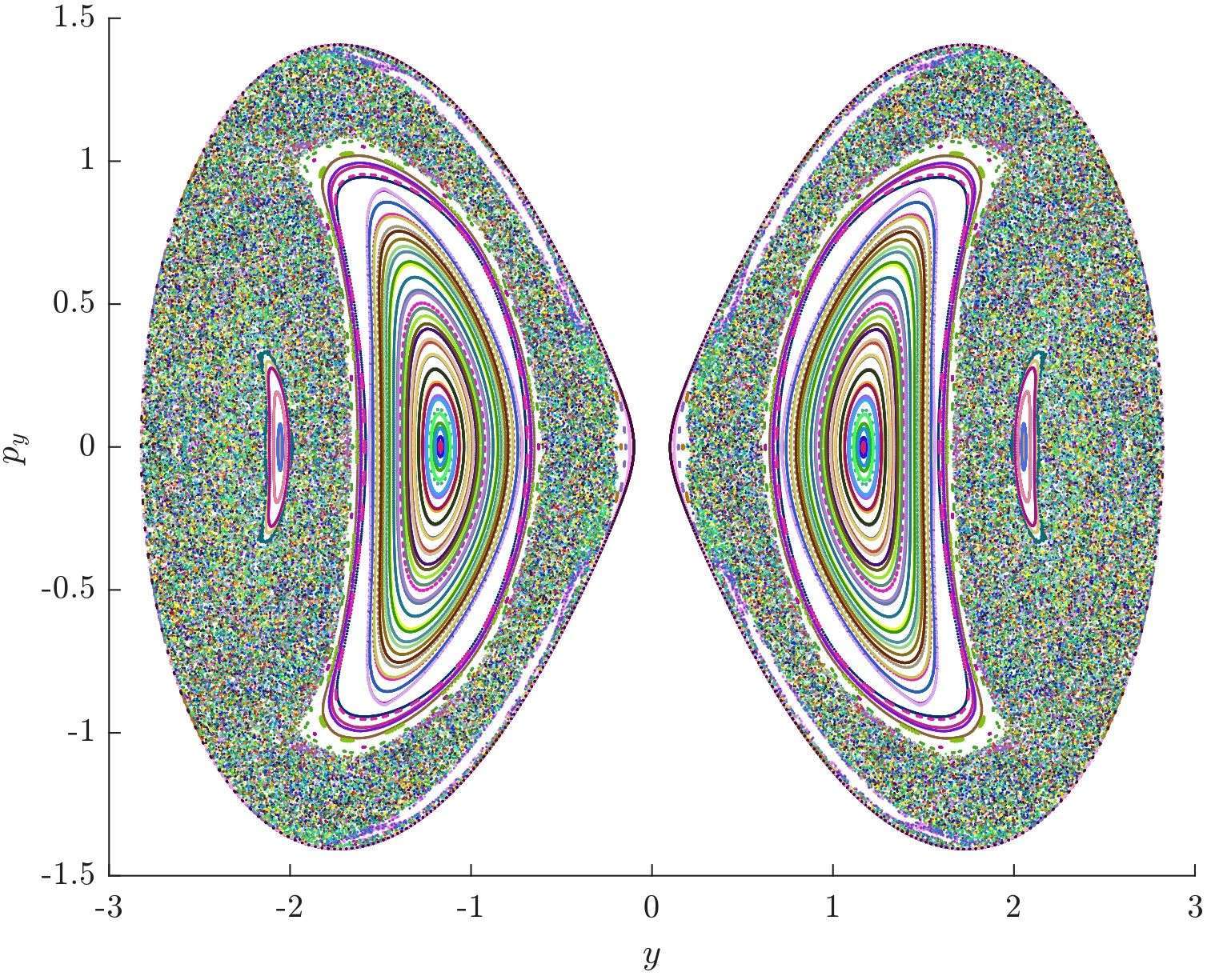}}
	\hfill
	\subfigure[$E=4\,E_s$]{\includegraphics[width=0.3\textwidth]{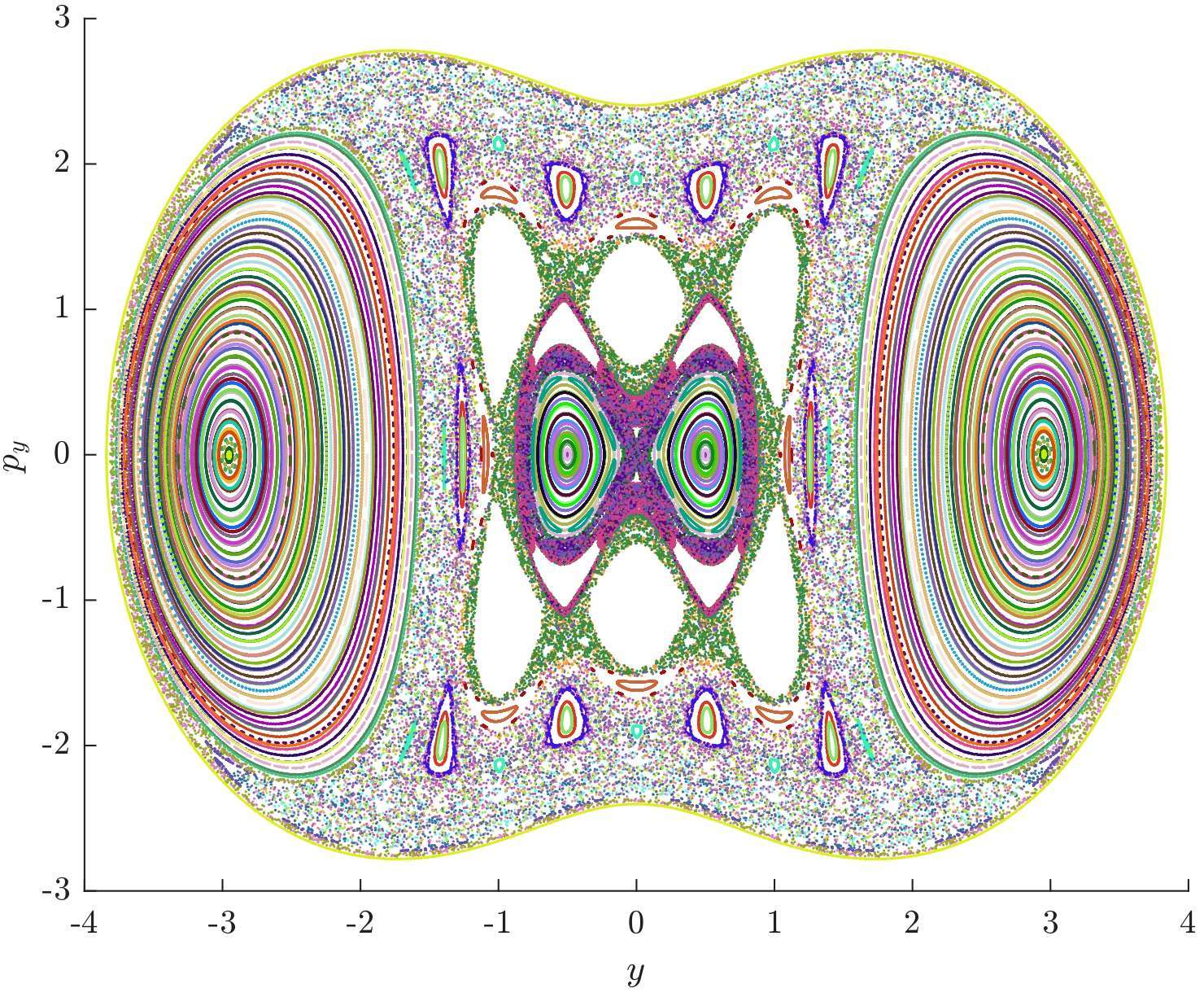}}
 \\
 \subfigure[$E=\frac{1}{2}E_s$]{\includegraphics[width=0.3\textwidth]{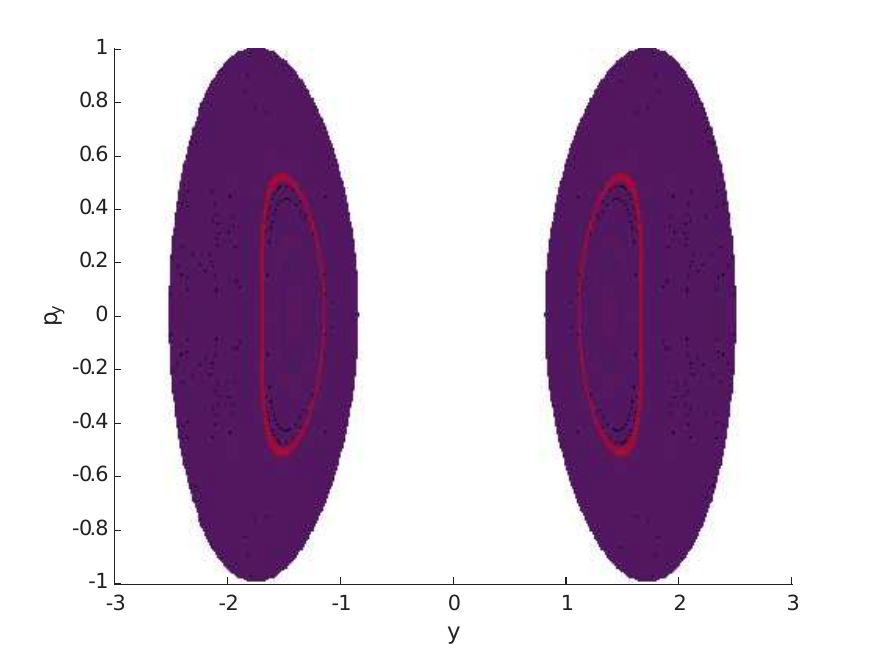}}
	\hfill
	\subfigure[$E=\frac{99}{100}E_s$]{\includegraphics[width=0.3\textwidth]{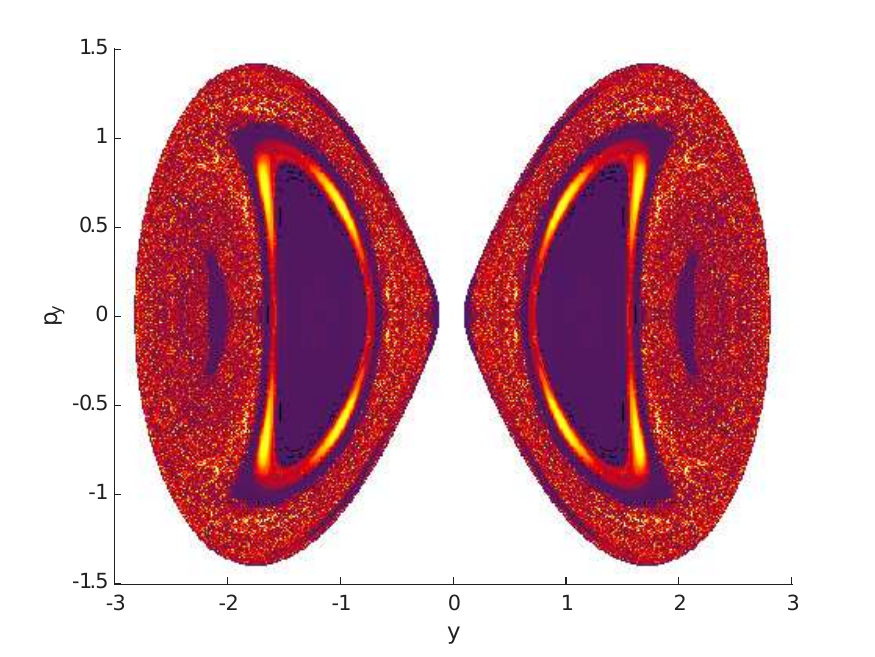}}
	\hfill
	\subfigure[$E=4\,E_s$]{\includegraphics[width=0.3\textwidth]{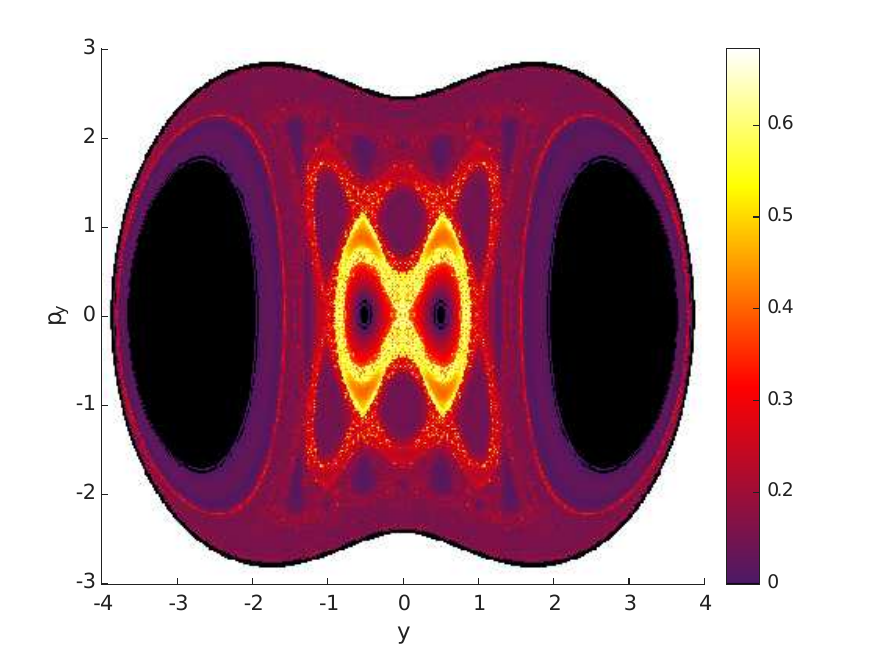}}
	\caption{\small (a)-(c) oriented Poincaré sections on the plane $(y,\,p_y)$, for the Hamiltonian $H$ (\ref{H1}) with $a=2$, at different values of the energy $H=E$; (d)-(f) the associated largest Lyapunov exponents. The equilibrium points are $(0,\,\pm \sqrt{3},\,0,\,0)$ and $E_s=1\,.$}
	\label{Fa2}
\end{figure*}

\begin{figure*}[h!]
	\centering
	\subfigure[$E=\frac{1}{2}E_s$]{\includegraphics[width=0.29\textwidth]{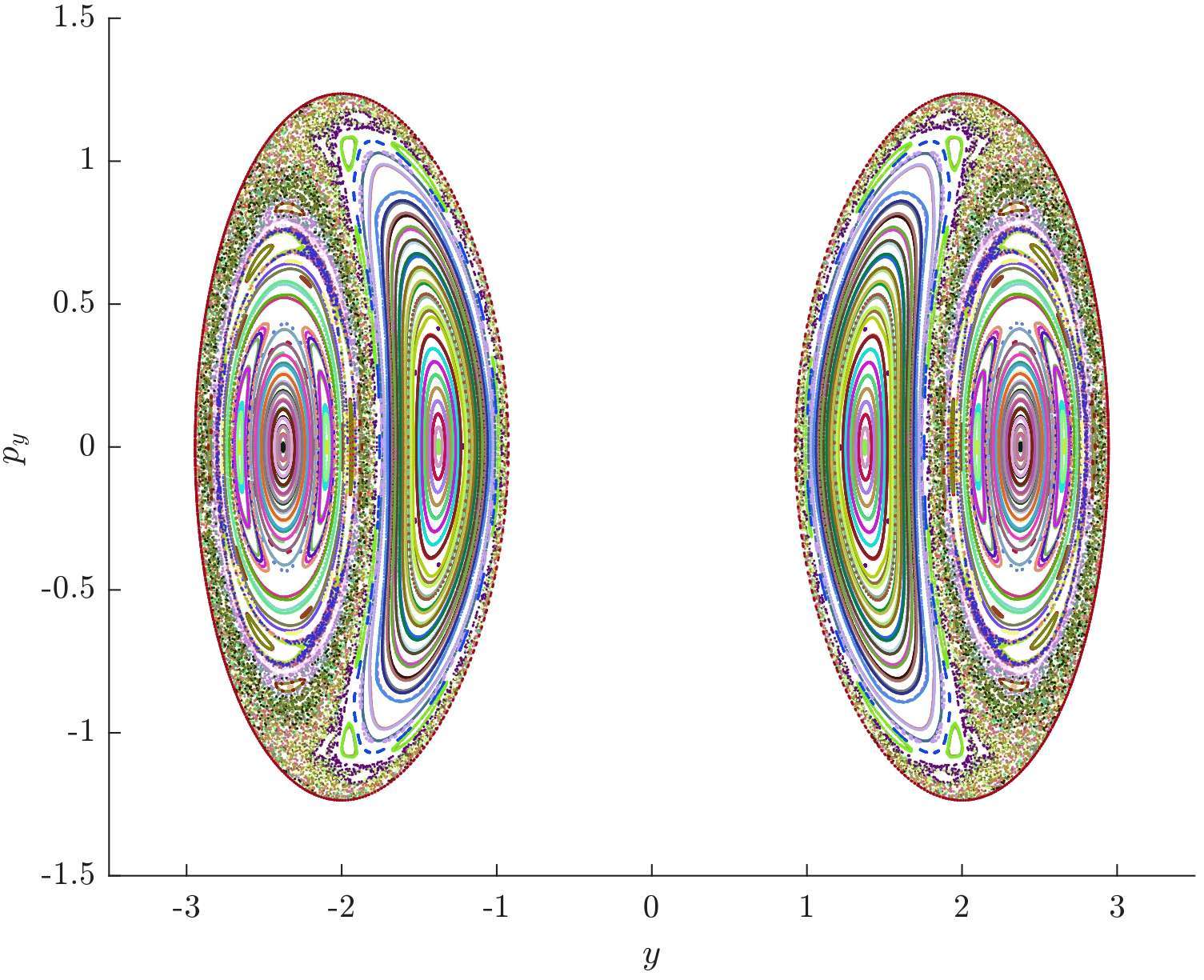}}
	\hfill 
	\subfigure[$E=\frac{99}{100}E_s$]{\includegraphics[width=0.29\textwidth]{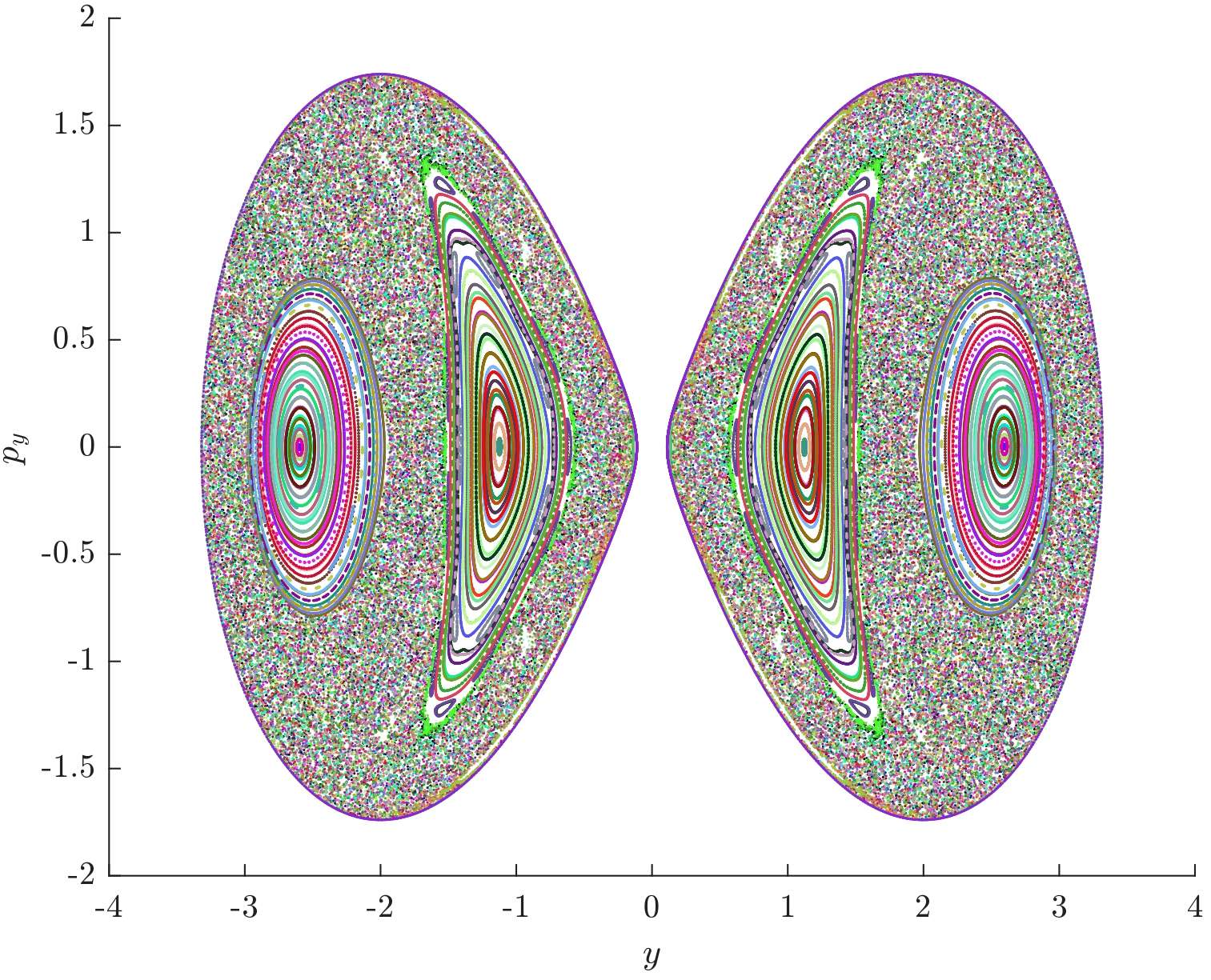}}
	\hfill
	\subfigure[$E=4\,E_s$]{\includegraphics[width=0.28\textwidth]{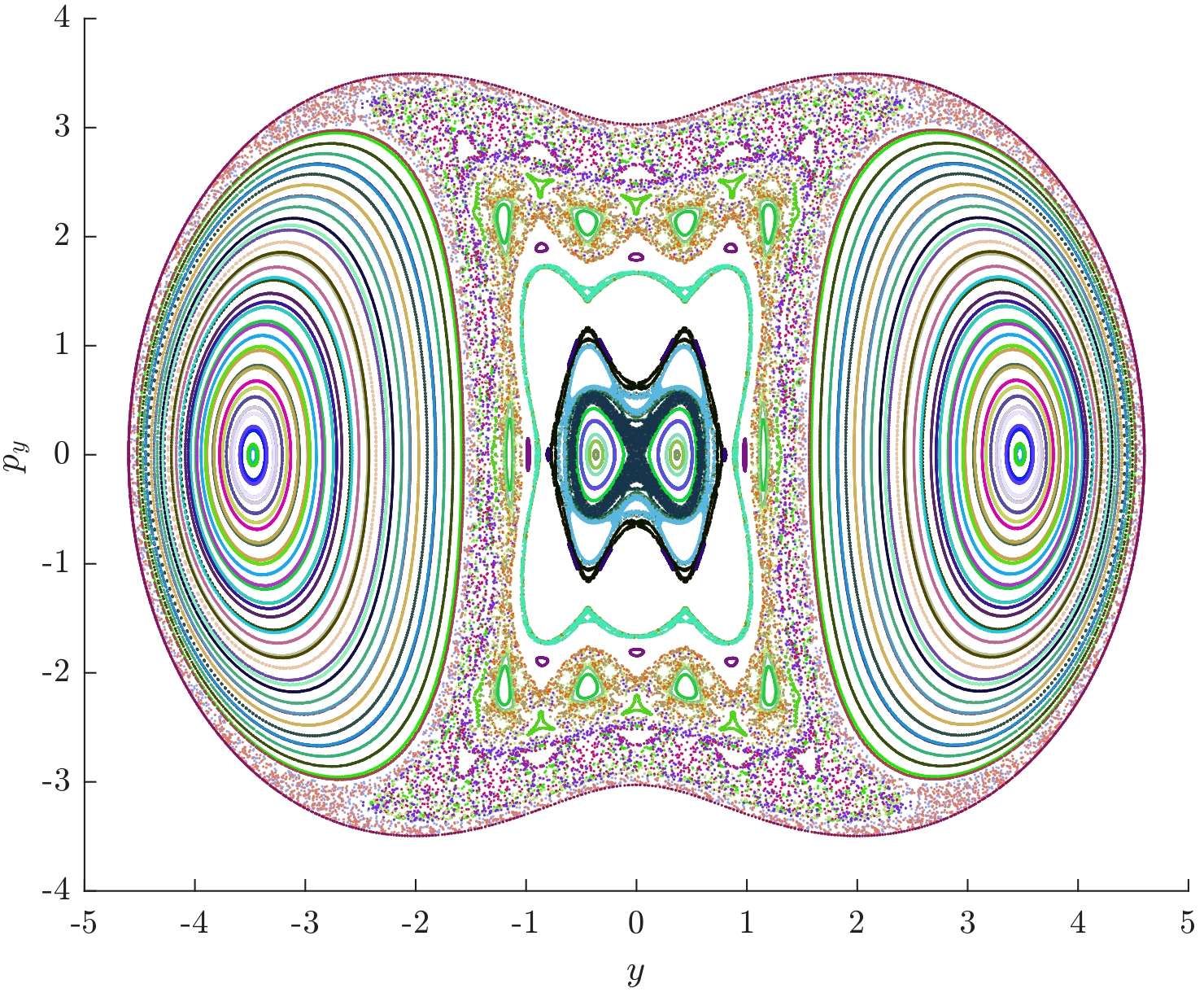}}
 \\ 
 \subfigure[$E=\frac{1}{2}E_s$]{\includegraphics[width=0.31\textwidth]{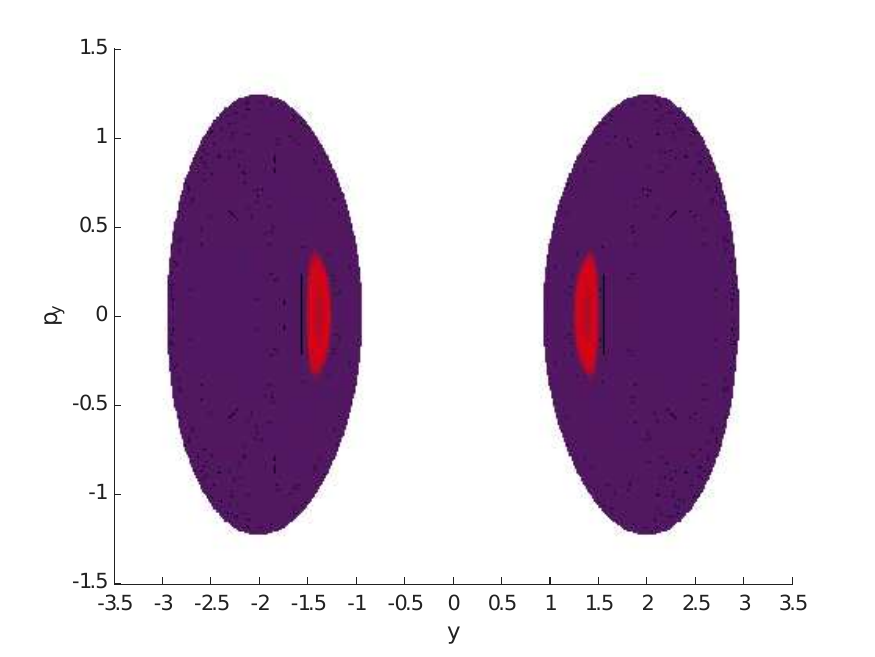}}
	\hfill
	\subfigure[$E=\frac{99}{100}E_s$]{\includegraphics[width=0.31\textwidth]{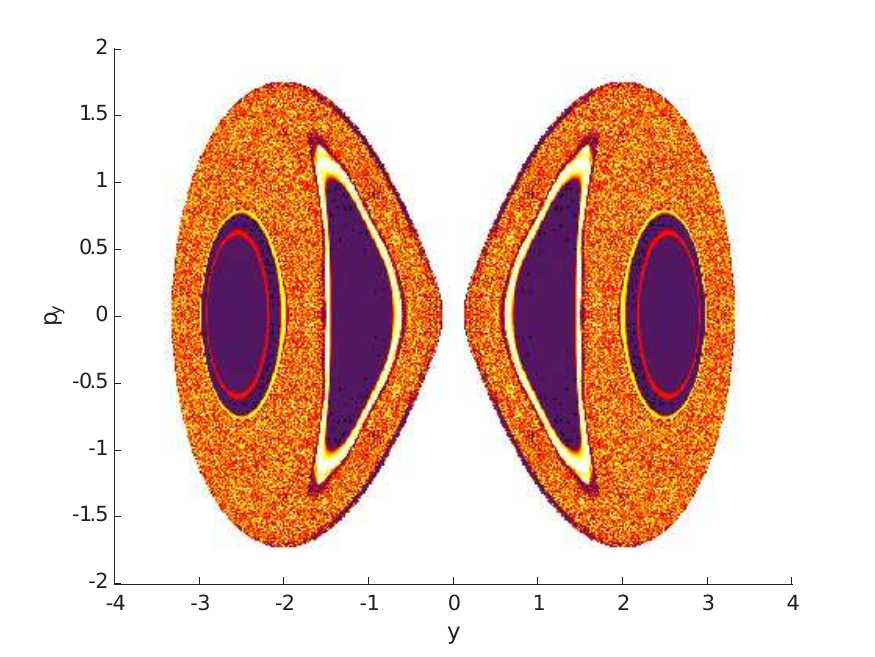}}
	\hfill
	\subfigure[ $\ E=4\,E_s$]{\includegraphics[width=0.31\textwidth]{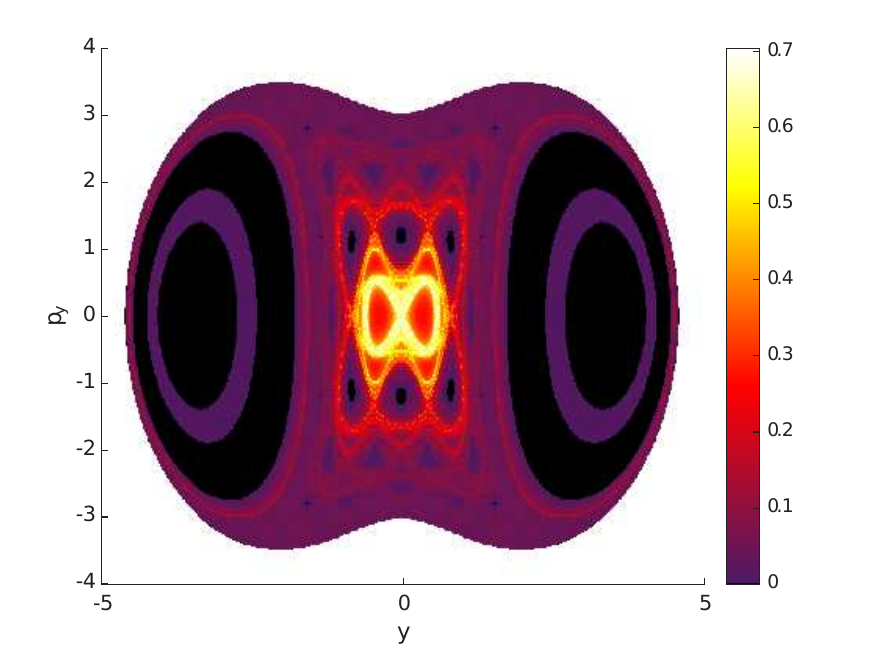}}
	\caption{\small (a)-(c) oriented Poincaré sections on the plane $(y,\,p_y)$, for the Hamiltonian $H$ (\ref{H1}) with $a=\sqrt{5}$, at different values of the energy $H=E$; (d)-(f) the largest Lyapunov exponents. The equilibrium points are $(0,\,\pm 2,\,0,\,0)$ and $E_s={(\sqrt{5}-1)}^2\,.$}
	\label{Fas5}
\end{figure*}

\begin{figure*}[h!]
\centering
\subfigure[$E= \frac{1}{2}E_s$]{\includegraphics[width=0.28\textwidth]{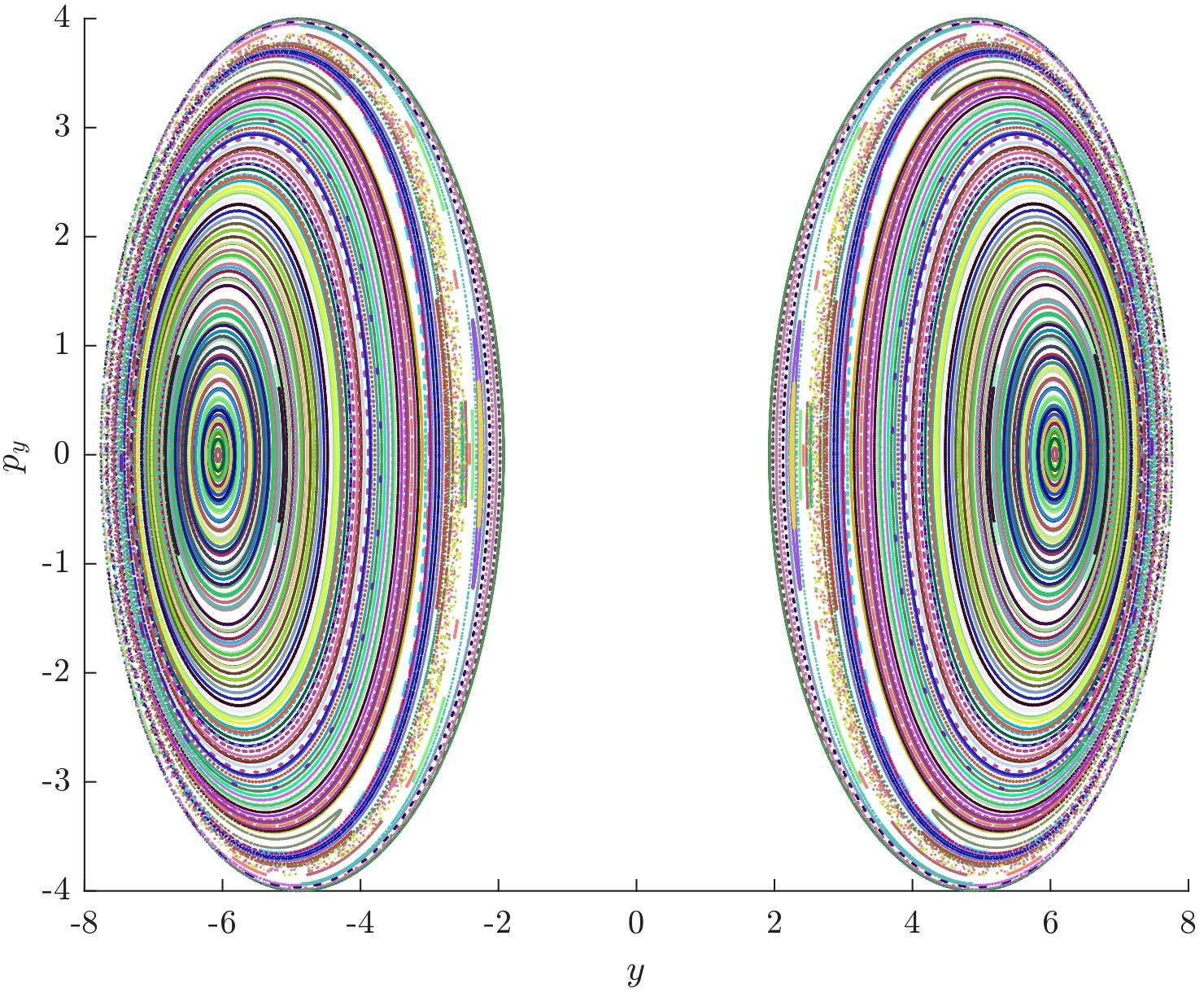}}
\hfill
\subfigure[$E= \frac{99}{100}E_s$]{\includegraphics[width=0.28\textwidth]{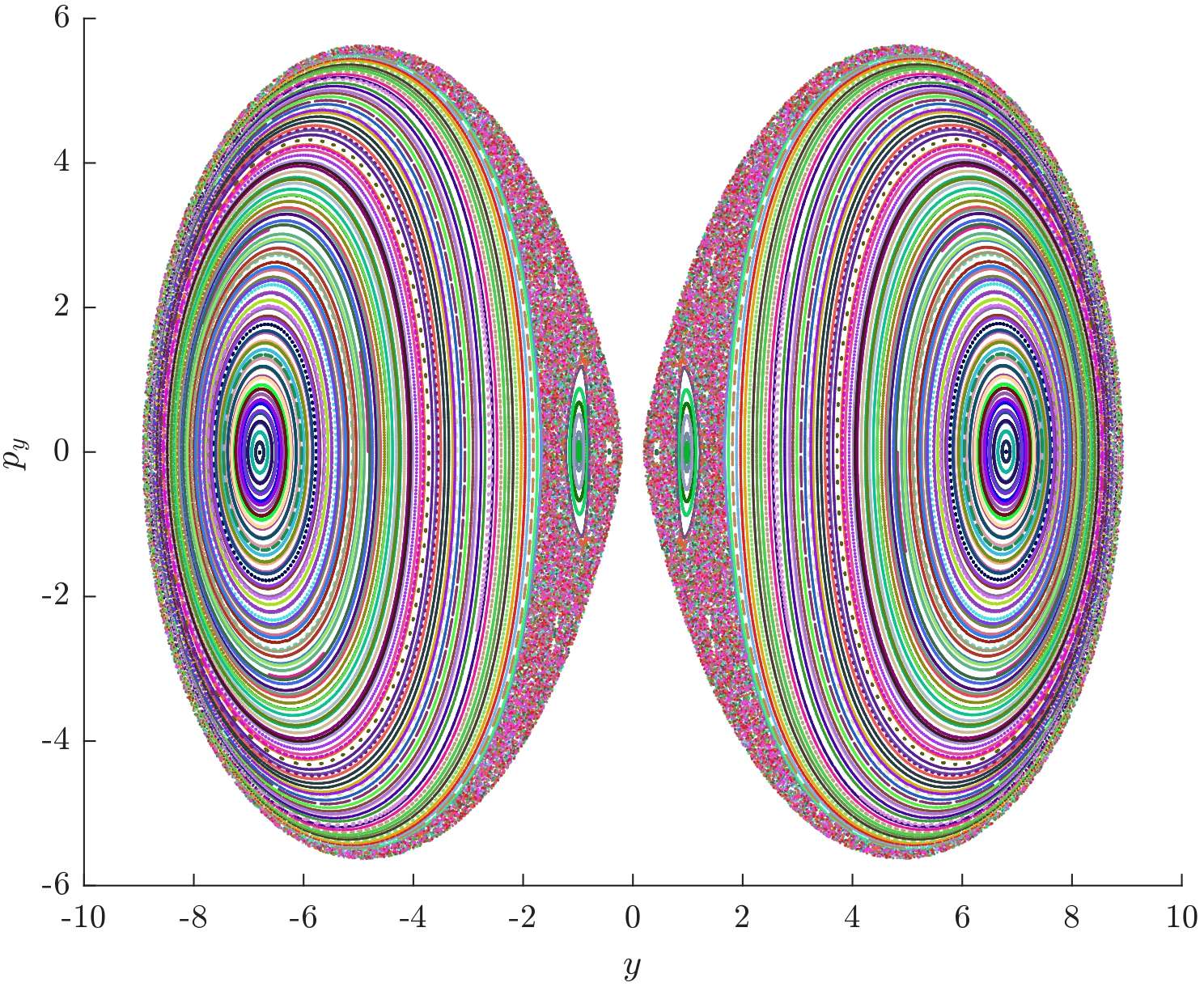}}
\hfill
\subfigure[$E= 4\,E_s$]{\includegraphics[width=0.3\textwidth]{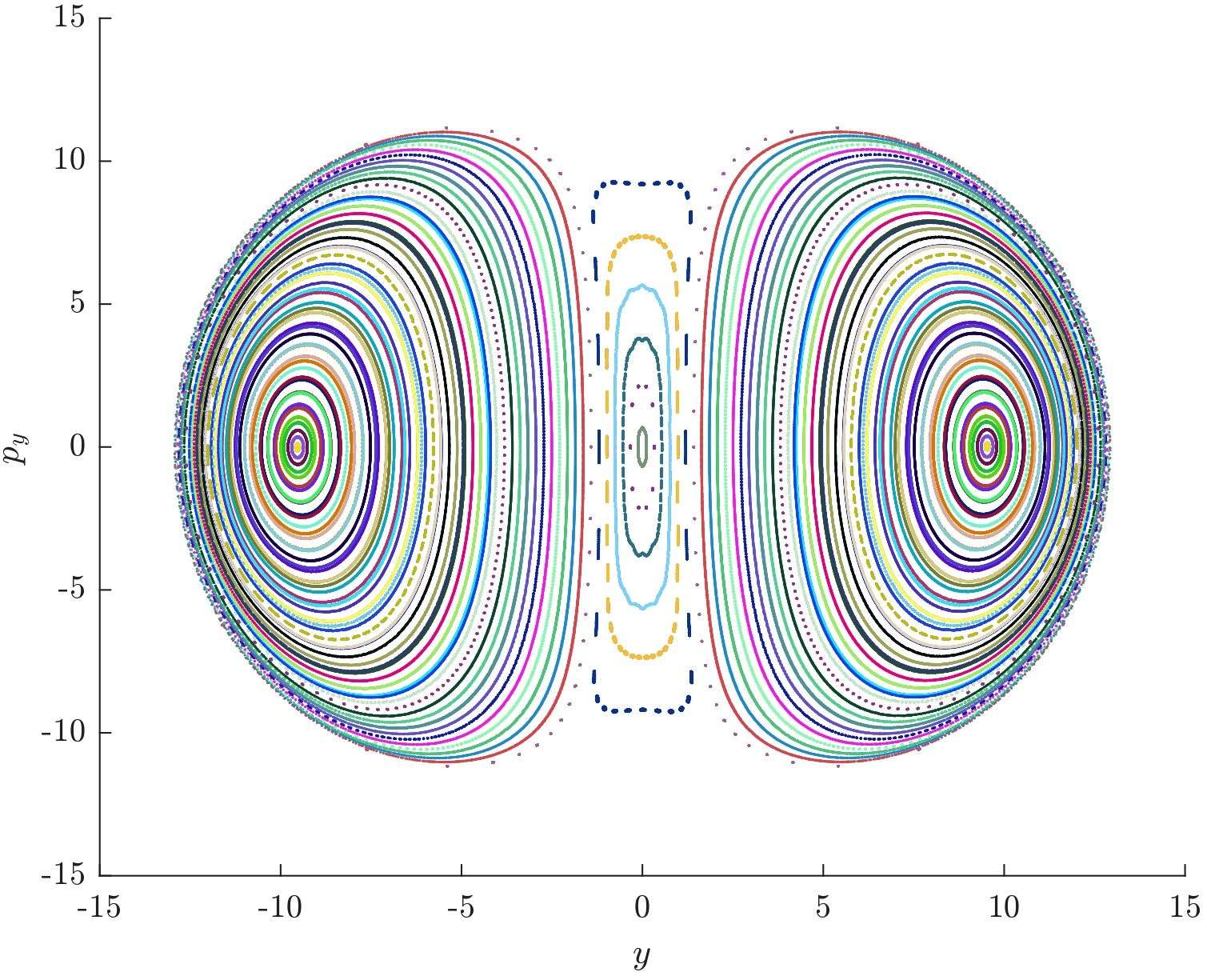}}
\\
\subfigure[$E= \frac{1}{2}E_s$]{\includegraphics[width=0.31\textwidth]{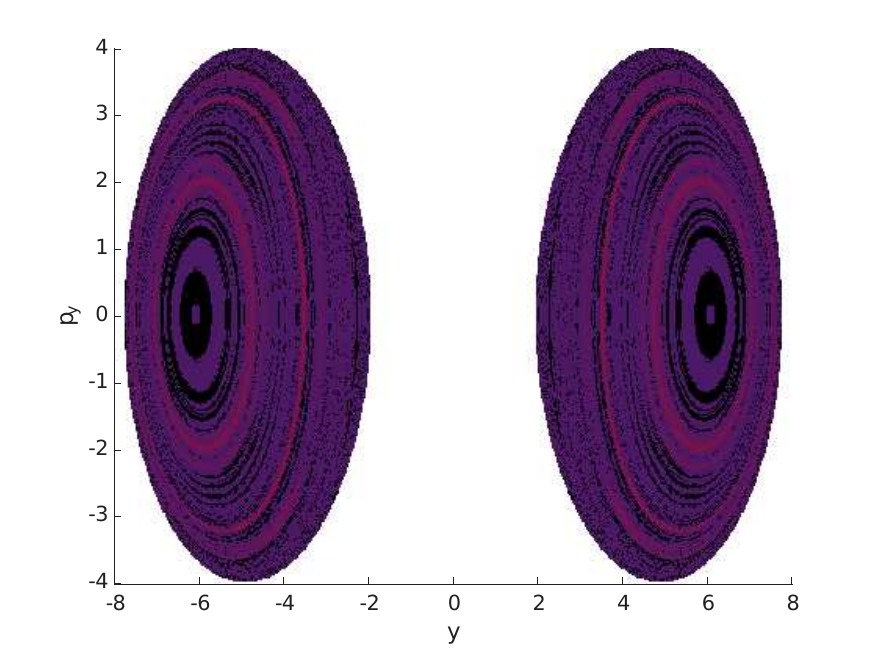}}
\hfill
\subfigure[$E= \frac{99}{100}E_s$]{\includegraphics[width=0.31\textwidth]{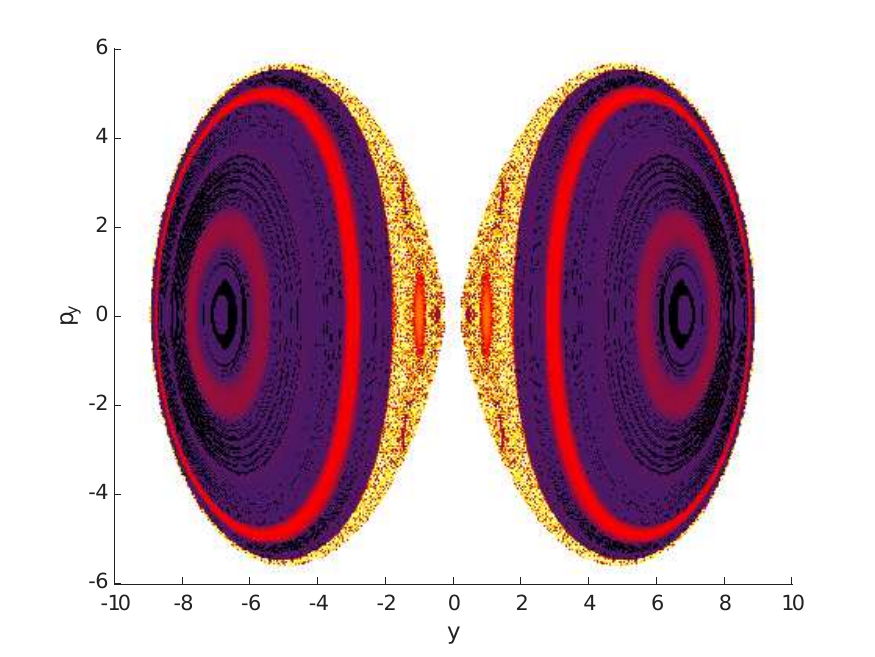}}
\hfill
\subfigure[$E= 4\,E_s$]{\includegraphics[width=0.32\textwidth]{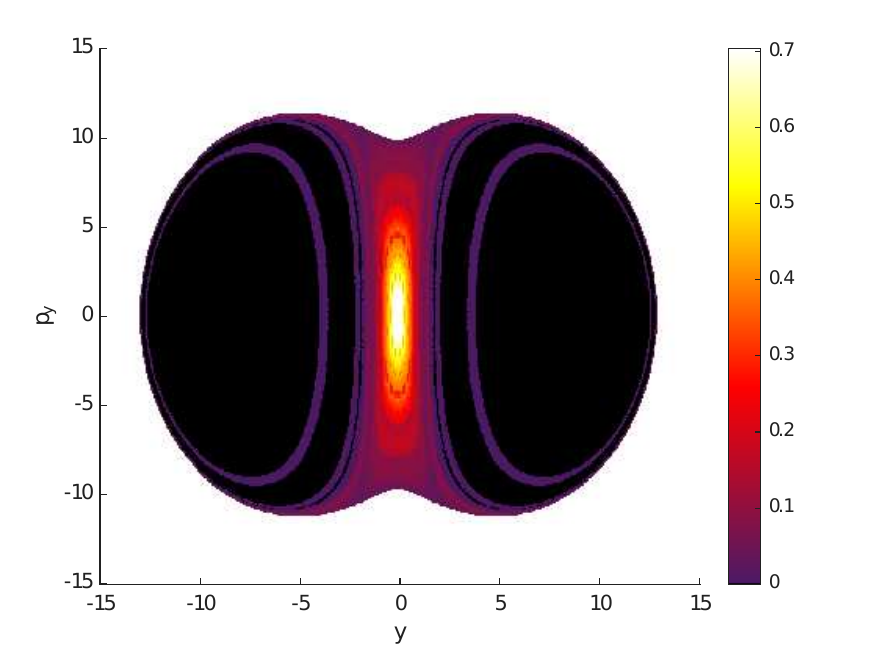}}
\caption{\small (a)-(c) oriented Poincaré sections on the plane $(y,\,p_y)$, for the Hamiltonian $H$ (\ref{H1}) with $a=5$, at different values of the energy $H=E$; (d)-(f) the largest Lyapunov exponents. The equilibrium points are $(0,\,\pm 2\,\sqrt{6},\,0,\,0)$ and $E_s=16\,.$}
\label{Fa5}
\end{figure*}

\clearpage

The top panel in Figures (\ref{Fa32}), (\ref{Fa2}), (\ref{Fas5}), and (\ref{Fa5}), displays the (oriented) Poincaré sections for the system (\ref{H1}) at $a=3/2,2,\sqrt{5},5$, respectively, as a function of the energy $E=H$ in units of $E_s$. They were obtained using numerical simulations for 120 random initial conditions with a simulation time of 6000.  Computations were performed in MATLAB utilizing a personal laptop. The corresponding maximum Lyapunov exponents are shown in the lower panel in Figures (\ref{Fa32}), (\ref{Fa2}), (\ref{Fas5}), and (\ref{Fa5}).

For $E<E_s$, the Poincaré sections consist of two symmetric disconnected regions which smoothly merge at $(y=0,p_y=0)$ when $E=E_s$. For low energy, $E \ll E_s$, the zones of regular dynamics dominate the accessible phase space landscape.

At fixed $a$, the presence of chaotic behaviour is prominent at $E\sim E_s$. Interestingly, at fixed energy $E$ (in units of $E_s$), additional numerical experiments indicate that the degree of chaoticity is not a monotonous function of the parameter $a$ in the interval $a \in (1,5]$ (see, for instance, the lower panel in Figures (\ref{Fa32}), (\ref{Fa2}), (\ref{Fas5}), and (\ref{Fa5}) for the case $E=\frac{99}{100}E_s$). Nevertheless, at fixed energy $E$ in units of $E_s$, the presence of chaos tends to decrease as the value of $a\gtrsim 3$ grows.

Notice that even at $E\sim E_s$, islands of stability (regular dynamics) persist. Hence, the coexistence of regularity and chaos exhibits the rich dynamics of the two-center problem with harmonic-like interactions. In particular, we highlight the complexity of the structure of the Poincaré sections for the case $a=\sqrt{5}$, see Fig. \ref{Fas5}.

\section{The averaging theory of first order}\label{s3}

It is worth recalling the basics of the averaging theory (periodic case) of first order. This tool will be used to derive the main results of the present study. Essentially, we deal with the problem of finding $T$-periodic solutions for a differential system whose vector field depends on a small parameter $\varepsilon$. For more details about the averaging theory of first order for finding periodic orbits see \cite{BL}.

We consider the differential system
\begin{equation}\label{xeps}
{\dot {\bf x}}(t)\ = \ \varepsilon  \,F_1(t,{\bf x}) \ + \ O(\varepsilon^2),
\end{equation}
where $\varepsilon \neq 0$ is a sufficiently small parameter, i.e. $|\varepsilon|\ll 1$, $F_1: \mathbb R\times \Omega \rightarrow  \mathbb R^n$ is a continuous function $T$-periodic in the variable $t$, and $\Omega$ denotes an open subset of $\mathbb R^n$. The above equation often arises by expansion in the neighborhood of an equilibrium point taking convenient coordinates.

Now we introduce the averaged function of first order $f_1:\Omega \rightarrow  \mathbb R^n$ as follows
\begin{equation}\label{f1av}
f_1({\bf z}) \ = \ \frac{1}{T}\int_0^T F_1(s,\,{\bf z})\,ds \ ,
\end{equation}
and also assume that:
\begin{itemize}
\item (i) $F_1$ is locally Lipschitz with respect to ${\bf x}$;
\item (ii) for ${\bf z}_0$ in $\Omega$ with $f_1({\bf z}_0)=0$, there exists a neighborhood $U$ of ${\bf z}_0$ such that $f_1({\bf z})\neq 0$ for all ${\bf z} \in U\setminus\{{\bf z}_0\}$ and $d_B(f_1,\,U,\,{\bf z}_0)\neq 0$  (the Brouwer degree of the $f_1$ at ${\bf z}_0$ is not zero).
\end{itemize}
Then for $|\varepsilon|$ sufficiently small, there exists a $T-$periodic solution ${\bf x}(t,\,\varepsilon)$ of system (\ref{xeps}) such that ${\bf x}(0,\,\varepsilon) \rightarrow {\bf z}_0$ as $\varepsilon \rightarrow 0$. That is, the simple zeros of the averaged function (\ref{f1av}) provide the initial conditions for isolated $T$-periodic solutions of the differential system (\ref{xeps}). Here a simple zero ${\bf z}_0$ of the function $f_1$ means that the Jacobian of $f_1$ at ${\bf z}_0$ is not zero.

We recall that if the Jacobian of $f_1$ at ${\bf z}_0$ is not zero, then the Brouwer degree $d_B(f_1,\,U,\,{\bf z}_0)\neq 0$, for details see \cite{Ll}.

\section{Proof of theorem \ref{t1}}\label{s4}

In this section we address the problem of finding periodic orbits of the differential system \eqref{dotx} bifurcating from the equilibrium point ${\mathbf x}_0 \,=\, (x,y,p_x,p_y) \ = \ (0,\,\sqrt{a^2-1},\,0,\,0)$, with $a>1$.

As a first step we translate the equilibrium ${\mathbf x}_0$ to the origin of coordinates. To this end, we introduce the canonical transformation
\begin{equation}\label{trasl}
(x,y,p_x,p_y)=(X,Y+g,P,Q).
\end{equation}
In these new variables the Hamiltonian (\ref{H1}) writes
\begin{equation}\label{Ht}
\begin{array}{rl}
H=&   \frac{1}{2}\big(\,P^2 \ + \ Q^2\,\big) \ + \ X^2 \ + \ Y^2 \ + \ 2\,a^2 \ + \ 2\,Y\,g \\ 
&- \ a \,\big(\sqrt{\left(g+Y\right)^2+(X-1)^2}\ + \  \sqrt{\left(g+Y\right)^2+(X+1)^2}\big), \nonumber
\end{array}
\end{equation}
and its Hamilton's equations are
{\small
\begin{equation}\label{DS2}
\begin{aligned}
\dot{X}=& P, \\
\dot{Y}=& Q, \\
\dot{P}=& -\frac{(X-1) \left(\sqrt{(g+Y)^2+(X-1)^2}-a\right)} {\sqrt{(g+Y)^2+(X-1)^2}}\\
& -\frac{(X+1) \left(\sqrt{(g+Y)^2+(X+1)^2}-a\right)} {\sqrt{(g+Y)^2+(X+1)^2}},\\
\dot{Q}=&  \ (g+Y) \bigg(a \bigg(\frac{1}{\sqrt{(g+Y)^2+(X+1)^2}}
\\ &
+ \frac{1} {\sqrt{(g+Y)^2+(X-1)^2}}\bigg)-2\bigg).
\end{aligned}
\end{equation}
}
Now it is convenient to perform another change of variables such that the linear part at the origin of the differential system (\ref{DS2}) be in its real Jordan normal form. Direct calculations lead to the transformations
\begin{equation}
\label{T2}
(X,Y,P,Q)=\left(U, W, -\frac{\sqrt{2}}{a}\,V, -\frac{\sqrt{2}\,g}{a} Z\right).
\end{equation}
From (\ref{DS2}) and (\ref{T2}) we obtain the differential system
\begin{equation}\label{DS3}
\begin{aligned}
\dot{U}&= -\frac{\sqrt{2}}{a}\, V, \\
\dot{W}&= -\frac{\sqrt{2} \,g}{a}\, Z,\\
\dot{V}&= \ -\frac{a}{\sqrt{2}} \bigg(\bigg[\frac{a\,(U+1)} {\sqrt{(g+W)^2+(U+1)^2}}
\\ &
+ \frac{a\,(U-1)}{\sqrt{(g+W)^2+(U-1)^2}}\bigg] - \ 2\,U\,\bigg), \\
\dot{Z}&= \frac{a\, (g+W)}{\sqrt{2} g} \, \bigg(2-a \bigg[\frac{1} {\sqrt{(g+W)^2+(U+1)^2}}
\\ &
+\frac{1}{\sqrt{(g+W)^2+(U-1)^2}}\bigg]\bigg).
\end{aligned}
\end{equation}
which admits from (\ref{Ht}) the first integral
{\small
\begin{equation}\label{HV}
\begin{array}{rl}
&H=  \dfrac{(V-Z) (V+Z)}{a^2}+a^2 + (g+W)^2 +U^2+Z^2+ 1\vspace{0.2cm}\\
&-a \left( \sqrt{(g+W)^2+(U-1)^2}+\sqrt{(g+W)^2+(U+1)^2}\right).
\end{array}
\end{equation}
}
Next we rescale the variables for rewriting system (\ref{DS3}) into a suitable form for applying the averaging theory. Let $\varepsilon$ be a small parameter, and we do the rescaling
\begin{equation}\label{scalee}
(U,W,V,Z)=(\varepsilon\,u, \varepsilon\,w,\varepsilon\,v,\varepsilon\,z),
\end{equation}
in the differential system (\ref{DS3}), and expanding the new differential system in powers of the small parameter $\varepsilon$ we obtain
\begin{equation}\label{DS4}
\begin{aligned}
    \dot{u}\ &= \ -\frac{\sqrt{2} \,v}{a}, \\
    \dot{w}\ &= \  -\frac{\sqrt{2}\, g\, z}{a}, \\
    \dot{v} \ &= \ \frac{\sqrt{2} \,u}{a}+\frac{\sqrt{2} \left(a^2-3\right)\, g\, u\, w}{a^3}\varepsilon + O(\varepsilon^2), \\
    \dot{z}\ &= \ \frac{\sqrt{2} \,g\, w}{a}
 \ + \ \frac{ \left(\left(a^2-3\right) u^2+3 w^2\right)}{\sqrt{2} \,a^3}\,\varepsilon + O(\varepsilon^2).
\end{aligned}
\end{equation}
whereas for (\ref{HV}) the first integral becomes
\begin{equation*}
H \ = \ \frac{ \left(g^2 \left(w^2+z^2\right)+u^2+v^2\right)}{a^2}\varepsilon ^2 \ +  \ O(\varepsilon^3).
\end{equation*}

We introduce the following polar variables $(r,\theta,\rho,\,s)$
\begin{equation}\label{polarcor}
\begin{array}{ll}
u \,=\, r\,\cos(\sqrt{2}\,\theta / a),  & v \,=\, r\,\sin(\sqrt{2}\,\theta / a),\\
w \,=\, \rho\,\cos(\sqrt{2}\,g\,(\theta+s) / a), &  z \,=\, \rho\,\sin(\sqrt{2}\,g\,(\theta +s)/ a).
\end{array}
\end{equation}
Now we take the angular variable $\theta$ as the new independent variable. So, from (\ref{DS4}) and (\ref{polarcor}) we arrive to the differential system
\begin{equation}\label{DS5}
\begin{array}{rl}
\dfrac{dr}{d\theta}=& \dfrac{\varepsilon}{\sqrt{2}a^3}\left(a^2-3\right)\, g\, \rho\,  r \, \sin \frac{2 \sqrt{2} \theta }{a} \cos \frac{\sqrt{2} g (\theta +s)}{a}+ O(\varepsilon^2), \vspace{0.2cm}\\
\dfrac{d\rho}{d\theta}=& \dfrac{\varepsilon}{\sqrt{2}a^3} \sin \frac{\sqrt{2}g (\theta +s)}{a} \bigg(\left(a^2-3\right)r^2 \cos ^2\frac{\sqrt{2} \theta }{a}
\\ &
+ \ 3\, \rho ^2 \cos ^2\frac{\sqrt{2} g (\theta +s)}{a}\bigg)\ + \  O(\varepsilon^2), \vspace{0.2cm}\\
\dfrac{ds}{d\theta}=& \dfrac{\varepsilon}{4a^2g\rho} \cos\frac{\sqrt{2} g (\theta +s)}{a} \bigg(\left(a^2-3\right) \cos\frac{2 \sqrt{2} \theta }{a} \left(r^2-2 g^2 \rho ^2\right)
\\ &
+ \ \left(a^2-3\right) r^2  \ + \ \left(8 a^2-2 a^4-3\right) \rho ^2
\\ &
+ \ 3 \,\rho ^2 \cos\frac{2 \sqrt{2} g (\theta +s)}{a}\bigg) \ + \   O(\varepsilon^2),
\end{array}
\end{equation}
possessing the first integral
\begin{eqnarray}
  &  H \ = \ \dfrac{ \left(g^2\,\rho ^2 \,+\,r^2\right)}{a^2}\,\varepsilon ^2 \ + \ O(\varepsilon^3). \nonumber
\end{eqnarray}

Since  in  the  Hamiltonian  systems  generically  the  periodic  orbits  appear  in  cylinders  of  periodic  orbits parametrized  by the  values  of  the  Hamiltonian  $H$, and  the  averaging  theory  only  can  detect  periodic  orbits  that  are  isolated, we  restrict  the  above differential  system (\ref{DS5})  to  the  energy  level  $H = \varepsilon^2\, h$  with $h > 0$.  We impose this restriction computing $r$ as a function of $\rho$ and $s$ in the energy level $H = \varepsilon^2\, h$, namely
\begin{equation*}\label{er}
r \ = \ \sqrt{a^2 \left(h-\rho ^2\right)\ + \ \rho ^2} + O(\varepsilon)\ .
\end{equation*}
Therefore, up to first order in $\varepsilon$ we obtain from (\ref{DS5}) the differential system
{\small
\begin{equation}\label{DS6}
\begin{array}{rl}
 \dfrac{d\rho}{d\theta}&= \dfrac{\varepsilon}{\sqrt{2} a^3}\sin \frac{\sqrt{2} g (\theta +s)}{a} \bigg(\left(a^2-3\right) \left(a^2 \left(h-\rho ^2\right)+\rho ^2\right) \cos ^2\frac{\sqrt{2} \theta }{a} \vspace{0.2cm} \\
& + \ 3 \rho ^2 \cos ^2\frac{\sqrt{2} g (\theta +s)}{a} \bigg) +O(\varepsilon^2)\vspace{0.2cm} \\
&= F_{11}(\theta,\rho,s)+O(\varepsilon^2), \vspace{0.2cm} \\
\dfrac{ds}{d\theta}&= \dfrac{\varepsilon}{4a^2g\rho} \cos \frac{\sqrt{2} g (\theta +s)}{a}\bigg(\left(a^2-3\right) \left(a^2 \left(h-3 \rho ^2\right)+3 \rho ^2\right) \cos\frac{2 \sqrt{2} \theta }{a}  \vspace{0.2cm} \\
&  +\left(a^2-3\right) a^2 h-3 \left(a^4-4 a^2+2\right) \rho ^2
\\ &
+\ 3 \rho ^2 \cos \frac{2 \sqrt{2} g (\theta +s)}{a}\bigg)+O(\varepsilon^2)\vspace{0.2cm} \\
&= F_{12}(\theta,\rho,s)+O(\varepsilon^2).
\end{array}
\end{equation}
}

The differential systems \eqref{DS3}, \eqref{DS4}, \eqref{DS5}, and \eqref{DS6}—the last two of which incorporate the terms of order $\epsilon^2$ not explicitly written—are equivalent, that is, they represent the same differential system expressed in different variables. A key advantage of first-order averaging theory, when applicable (i.e., when its assumptions are met, like in our case), is that terms up to first order in $\epsilon$ are sufficient to determine the existence of periodic orbits for the systems \eqref{DS3}, \eqref{DS4}, \eqref{DS5}, and \eqref{DS6}.

By direct integration we compute the first averaged function $f(\rho,s)=(f_1(\rho,s),f_2(\rho,s))$ with $T=\sqrt{2}\,a\,\pi$, i.e.
{\small
\begin{equation}\label{f}
\begin{array}{rl}
f_1=& \dfrac{1}{T}\displaystyle\int_0^{T} F_{11}(\theta,\rho,s)\, d\theta \vspace{0.2cm} \\
=& \dfrac{\sin \left(\pi  \sqrt{a^2-1}\right) \sin \left(\frac{\sqrt{a^2-1} \left(\pi  a+\sqrt{2} s\right)}{a}\right)}{2 \sqrt{2} \pi  a^3 \left(a^2-5\right) \sqrt{a^2-1}} \Bigg[ \
 2\, a^2\, \left(a^2-3\right)^2\, h
\\ & \hspace{-0.5cm}
+  \left(a^2-5\right) \rho ^2 \bigg(\cos \left(\frac{2 \sqrt{2} \sqrt{a^2-1} s}{a}\right)  +  \cos \left(\frac{2 \sqrt{a^2-1} \left(\pi  a+\sqrt{2} s\right)}{a}\right)
\\ & \hspace{-0.5cm}
+  \cos \left(\frac{2 \sqrt{a^2-1} \left(2 \pi  a+\sqrt{2} s\right)}{a}\right)+\cos \left(2 \pi  \sqrt{a^2-1}\right)\bigg)
\\ & \hspace{-0.5cm}
-  2 \left(a^6-7 a^4+14 a^2-4\right) \rho ^2
\Bigg]\ , \vspace{0.2cm} 
\end{array}
\end{equation}
\begin{equation*}
\begin{array}{rl}
f_2=&\dfrac{1}{T}\displaystyle\int_0^{T} F_{12}(\theta,\rho,s)\, d\theta \vspace{0.2cm} \\
=&  \dfrac{\sin \left(\pi  \sqrt{a^2-1}\right)\,\cos \left(\frac{\sqrt{a^2-1} \left(\pi  a+\sqrt{2} s\right)}{a}\right)}{4\, \pi\,  a^2 \left(a^2-5\right)\left(a^2-1\right)  \rho}   \Bigg[ 
2 a^2 \left(a^2-3\right)^2 h
\\ & \hspace{-0.5cm}
+ \ \left(a^2-5\right) \rho ^2 \bigg(\cos \left(\frac{2 \sqrt{2} \sqrt{a^2-1} s}{a}\right)  +  \cos \left(\frac{2 \sqrt{a^2-1} \left(\pi  a+\sqrt{2} s\right)}{a}\right)
\\ & \hspace{-0.5cm}
+ \ \cos \left(\frac{2 \sqrt{a^2-1} \left(2 \pi  a+\sqrt{2} s\right)}{a}\right)-\cos \left(2 \pi  \sqrt{a^2-1}\right)\bigg)
\\ & \hspace{-0.5cm}
+ \ 2 \left(-3 a^6+21 a^4-43 a^2+17\right)\, \rho ^2
\Bigg] \ .
\end{array}
\end{equation*}
}

The above expressions vanish for all $ \rho$ and $s$ when $a=\sqrt{N^2+1}$ with $N\in \Z$, then we assume $ a\neq  \sqrt{N^2+1}$.

We are interested in the zeros $(\tilde \rho,\tilde s)=(\tilde \rho(a,\,h), \tilde s(a,\,h))$ of the function $f(\rho,\,s)$. From the averaging theory described in section \ref{s3} such solutions must satisfy the requirements:
\begin{equation}\label{rderho}
\begin{aligned}
&\tilde \rho>0, \quad -2\,\pi \,\leq \tilde s<2\pi\ , \quad \tilde r=  \sqrt{a^2 (h-\tilde \rho ^2)+\tilde \rho ^2}>0\ , \\ & \dfrac{\partial(f_1,\,f_2)}{\partial(\rho,\, s)}\vert_{\rho=\tilde \rho,s=\tilde s}\neq 0\ .
\end{aligned}
\end{equation}

Since the variable $\rho^2$ appears in (\ref{f}) linearly, we solve $f_1(\rho,s)=0$ for $\rho^2$ and then substitute its expression into the equation $f_2(\rho,s)=0$. So, we arrive to the very simple solution
\begin{equation}
s \ = \ {\tilde s} \ \equiv \ \frac{\pi \, a\, (n-g)}{\sqrt{2} \,g}  
\end{equation}
which is $h-$independent, here $n$ is an integer number. In practice, only the two cases $n=0$ and $n=1$ will provide the relevant results. Substituting the above value $\tilde s$ into the equation  $f_1=0$ \eqref{f} we obtain $\tilde{\rho}$. For any integer $n$
{\small
\[
\tilde{\rho} =   a\,\sqrt{\frac{2\, \left(a^2-3\right)^2 h}{6 a^6-42 a^4+85 a^2-\left(a^2-5\right) \cos \left(2 \pi  \sqrt{a^2-1}\right)-29}}  \ .
\] 
}
Next, substituting this $\tilde{\rho}$ in equation \eqref{rderho} we obtain $\tilde{r}$
{\small
\[
\tilde{r}  =  a\,\sqrt{ h \left(1-\frac{2 \left(a^2-3\right)^2 \left(a^2-1\right)}{6 a^6-42 a^4+85 a^2-\left(a^2-5\right) \cos \left(2 \pi  \sqrt{a^2-1}\right)-29}\right)} \ .
\]
}
So, from the averaging theory of section \ref{s3}, we have the initial conditions for a periodic orbit of the differential system \eqref{DS6} if the determinant of the matrix
\begin{equation*}
A\ \equiv \ \left. \frac{\partial (f_1(\rho,s),f_2(\rho,s))}{\partial(\rho,s)}\right|_{\rho=\tilde \rho,\,s=\tilde s}\ne 0 \ .
\end{equation*}
A direct calculation leads to the result
{\small
\begin{equation}
\begin{aligned}    
{\rm Det}A \ & =   \ \frac{ (a^2-3)^2\,h\,\sin^2(\pi  \sqrt{a^2-1})  }{\pi^2 a^4 (a^2-5)^2 (a^2-1)} \bigg( (2 a^6-14 a^4+29 a^2
\\ &
+\ \left(a^2-5\right) \cos \left(2 \pi  \sqrt{a^2-1}\right)-13) \bigg)  \ .
\label{deter}
\end{aligned}
\end{equation}
}
We have already assumed that $a\neq\sqrt{N^2+1}$ for all integer $N$. Aditionally,  we assume  $a\neq\sqrt{3}$. The term in parentheses in \eqref{deter} is always non-zero since $a>1$.

Going back from polar coordinates \eqref{polarcor}, the scale factor $\varepsilon$ \eqref{scalee}, the change of coordinates \eqref{T2} and the translation transformations \eqref{trasl}, we obtain for $\varepsilon>0$ sufficiently small the initial conditions $(\,x(0,\varepsilon),y(0,\varepsilon),p_x(0,\varepsilon),p_y(0,\varepsilon))$ for a periodic orbit of the Hamiltonian system \eqref{dotx} in an energy level of $H=\tilde h>0$ sufficiently small because $\tilde h=h\,\varepsilon^2 +O(\varepsilon^3)$. More precisely, the initial conditions of such periodic orbit are
\begin{equation}\label{inicon}
\left( \varepsilon\,\tilde r,\ \varepsilon\,\tilde \rho\,\cos \frac{\sqrt{2} g} {a}\tilde s + \sqrt{a^2-1},\ 0,  \ -\,\varepsilon\dfrac{\sqrt{2}\, g}{a} \tilde \rho\, \sin \frac{\sqrt{2}\, g} {a} \tilde s\,\right),
\end{equation}
with an error of $O(\varepsilon^2)$. Theorem \ref{t1} is proved.

\section{Numerical examples}\label{s5}

In this section we present some examples of periodic orbits with initial conditions calculated by the averaging theory discussed in the previous section. The equilibrium point from which they bifurcate is $ {\bf x}_0 $, and their frequencies in the $ x$ and $ y$ directions are $ \omega_x = \sqrt{2}/a $ and $ \omega_y = \sqrt{2}g/a $, respectively. The frequency ratio of the periodic motion in the $(x,p_x)$ and $(y,p_y)$  planes,  $\omega_x/\omega_y=1/g=1/\sqrt{a^2-1}$,  is determined by the parameter $a$. Although we have found periodic solutions on both planes for $h>0$ sufficiently small, the motion in the phase space can be either periodic or quasi-periodic, depending on whether the frequency ratio $\omega_x/\omega_y=1/\sqrt{a^2-1}\ $ is a rational or an irrational number. If the commensurability condition $\omega_x/\omega_y= l/j\ \in \mathbb{Q}$ holds, with $l$, $j \,  \in \mathbb{Z}$  relative primes, the orbit will be periodic for $h>0$ sufficiently small, with $l$ oscillations in the $x$ direction and $j$ oscillations in the $y$ direction. In order to guarantee the periodicity of the solutions constructed from the averaging method, $a$ must be given by
\begin{equation*}
a\ = \ \sqrt{\left(\frac{j}{l}\right)^2\,+\,1}\,, \qquad l\,, j \in \mathbb{Z} \ .
\end{equation*}
Below we present explicit examples of periodic orbits obtained from the above described averaging method.

\vspace{-0.7cm}

\subsection{Case $a=\sqrt{13}/3$} 

For $h=1$, we obtain two zeros of the function (\ref{f}) such that $\dfrac{\partial(f_1,\,f_2)}{\partial(\rho,\, s)}\vert_{\rho=\tilde \rho,s=\tilde s}\,\neq \,0$, namely
\begin{equation*}
(\tilde \rho_1,\tilde s_1 )= (0.557978\,,\,-2.66984)\ ,
\end{equation*}
and
\begin{equation*}
(\tilde \rho_2,\tilde s_2 )= (0.557978\,,\,1.33492)\ .
\end{equation*}

\begin{figure}[htp!]
	\subfigure[$x$ \textit{vs} $p_x$]{\includegraphics[width=0.32
	\textwidth]{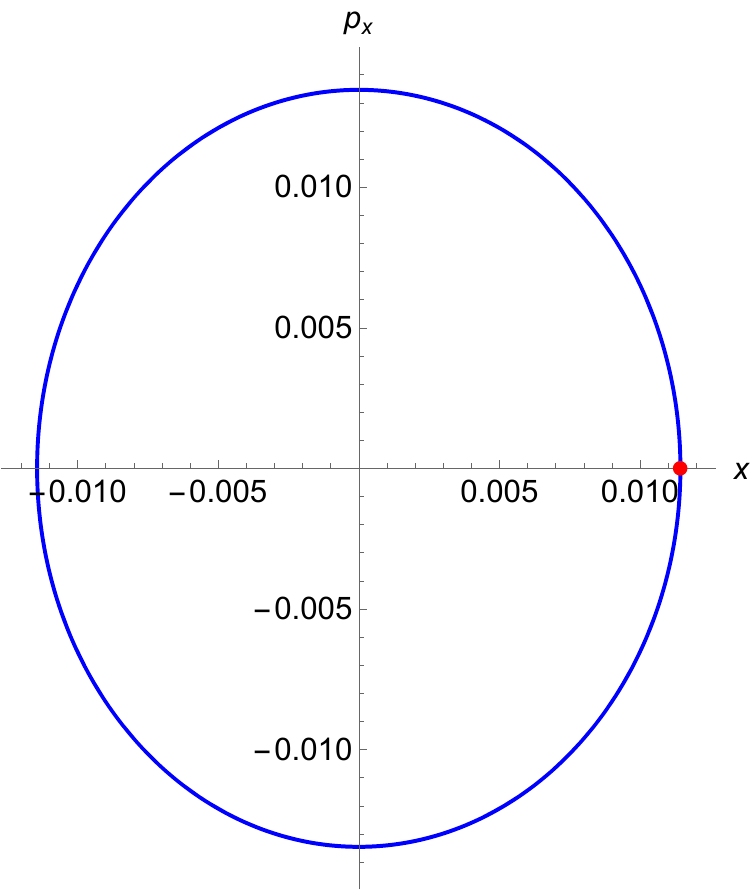}}
 \hfill
	\subfigure[$p_x$ \textit{vs} $p_y$]{\includegraphics[width=0.45
\textwidth]{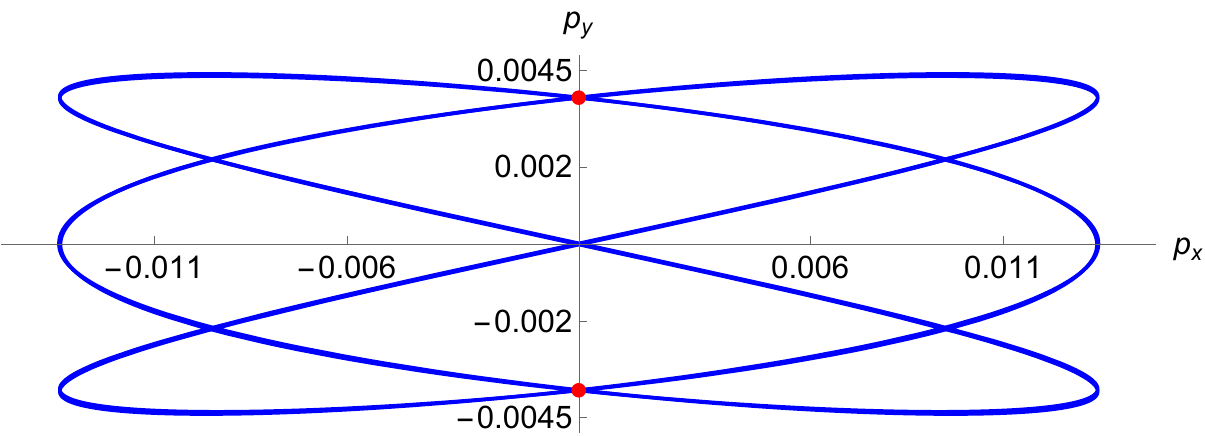}}
\hfill
	\subfigure[$x$ \textit{vs} $y$]{\includegraphics[width=0.41\textwidth]{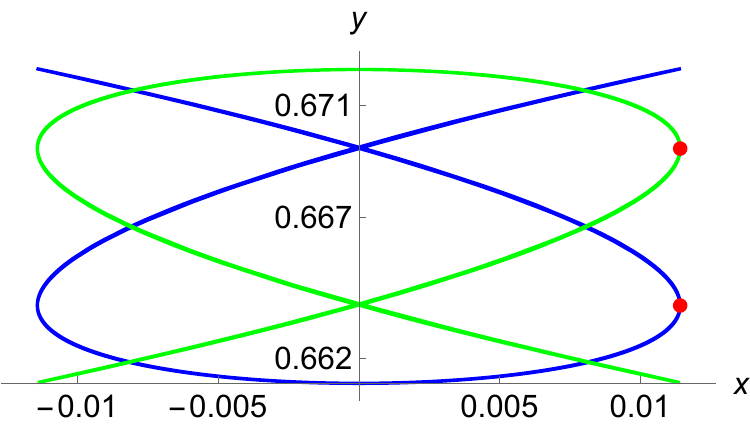}}
	\caption{Case $a=\sqrt{13}/3$. (a) and (b) periodic orbit and its initial conditions (\ref{inicondsqrt13a}) obtained using the averaging theory on the $(x,p_x)$  and $(p_x,p_y)$ planes, respectively. The initial conditions are marked by red dots. (c) The two periodic orbits that bifurcate from the equilibrium point $(0,\frac{2}{3},0,0)$. A simulation time of $20\,T$ was used in the computations.}
	\label{orbita1}
\end{figure}

By taking $\varepsilon= 10^{-2}$ the above values $(\tilde \rho_1,\tilde s_1 )$ provide the following initial conditions (\ref{inicon}):
{\small
\begin{equation}\label{inicondsqrt13a}
\begin{split}
&(\,x(\varepsilon), y(\varepsilon),p_x(\varepsilon), p_y(\varepsilon)\,)\ = 
\ (0.011428,\, 0.663877,\, 0,\, 0.00379)\ ,
\end{split}
\end{equation}
}
which correspond to the periodic orbit displayed in Fig. \ref{orbita1}. In this case $\omega_x/\omega_y=1/\sqrt{a^2-1}=2/3$, and the periodic orbit has 2 oscillations in the  $x$ direction by 3 oscillations in the $y$ direction. From another side, the values $(\tilde \rho_2,\tilde s_2 )$ provide the initial conditions (\ref{inicon}):
{\small
\begin{equation}\label{inicondsqrt13b}
\begin{split}
&(\,x(\varepsilon), y(\varepsilon),p_x(\varepsilon), p_y(\varepsilon)\,)\ = 
\ (0.011428,\,0.66946, \,0,\, -0.00379)\ .
\end{split}
\end{equation} 
}

\vspace{-1.1cm}

\subsection{Case $a=\sqrt{5}$} 

This case was excluded from the proof of Theorem 1, and we address this special value here. The averaged function (\ref{f})  reduces to
{\small
\begin{equation}\label{ff}
\begin{array}{rl}	
f_1(\rho,\,s)=&  \dfrac{1}{T}\displaystyle\int_0^{T} F_{11}(\theta,\rho,s)\, d\theta \vspace{0.2cm} \\
=& \frac{\left(5 h-4 \rho ^2\right) \sin \left(2 \sqrt{\frac{2}{5}} s\right)}{10 \sqrt{10}} \ ,\vspace{0.2cm} 
\end{array}
\end{equation}
\begin{equation*}
\begin{array}{rl}
f_2(\rho,\,s)=&  \dfrac{1}{T}\displaystyle\int_0^{T} F_{12}(\theta,\rho,s) \,d\theta \vspace{0.2cm} \\
=& \frac{\left(5 h-12 \rho ^2\right) \cos \left(2 \sqrt{\frac{2}{5}} s\right)}{40 \rho } \ .
\end{array}
\end{equation*}
}

\begin{figure}[htp!]
	\centering
	\subfigure[$x$ \textit{vs} $p_x$]{\includegraphics[width=0.46
	\textwidth]{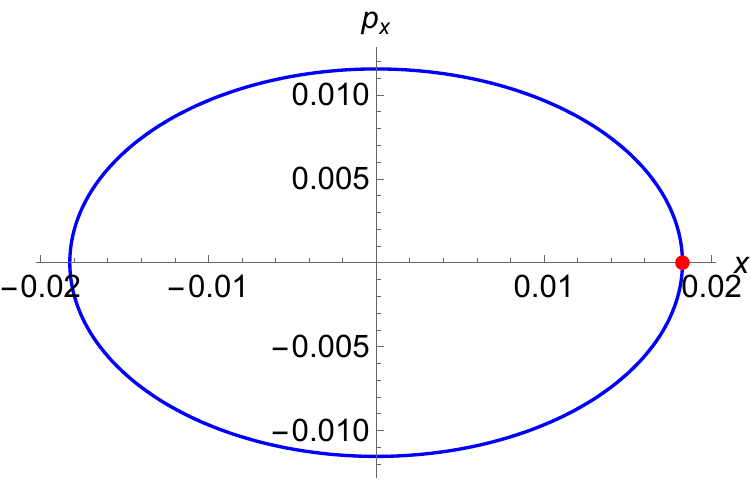}}
	\hfill
	\subfigure[$p_x$ \textit{vs} $p_y$]{\includegraphics[width=0.4
\textwidth]{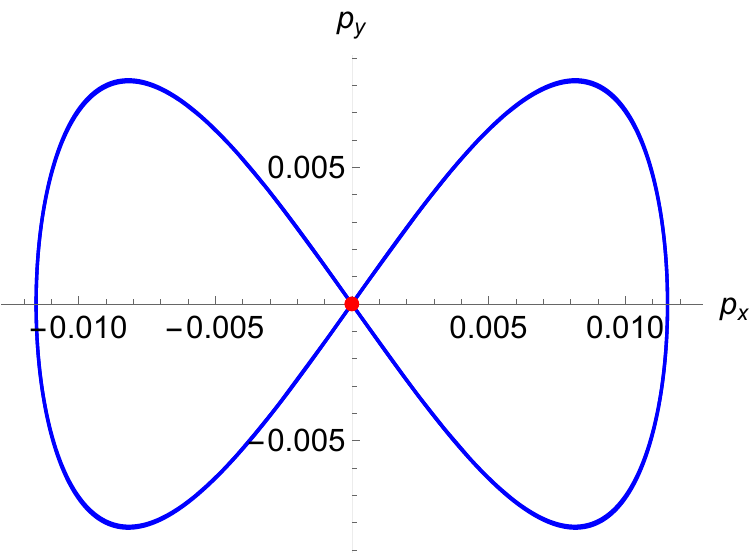}}
	\hfill
	\subfigure[$x$ \textit{vs} $y$]{\includegraphics[width=0.44\textwidth]{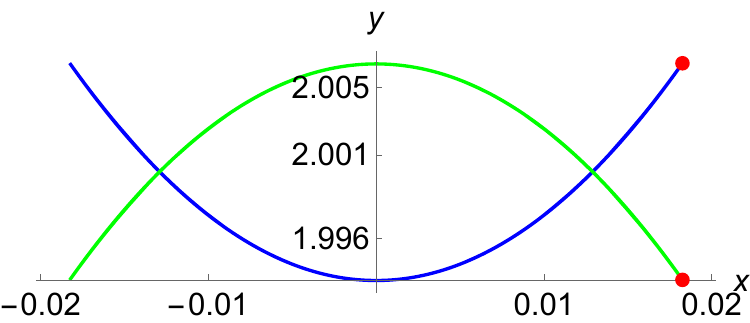}}
	\caption{Case $a=\sqrt{5}$. (a) and (b) periodic orbit and its initial conditions (\ref{inicondsqrt5}) obtained using the averaging theory on the $(x,p_x)$  and $(p_x,p_y)$ planes, respectively. The initial conditions are marked by red dots. (c) The two periodic orbits that bifurcate from the equilibrium point $(0,2,0,0)$. A simulation time of $20\,T$ was used.}
	\label{orbita2}
\end{figure}

For $a=\sqrt{5}$ and $h=1$ we found the zeros
\begin{equation*}
(\tilde \rho_1,\tilde s_1) \approx (0.645497,\,	-4.96729) \ ,
\end{equation*}
and
\begin{equation*}
(\tilde \rho_2,\tilde s_2)\ \approx (0.645497,\,	-2.48365) \ ,
\end{equation*}
of the averaged function $f=0$ (\ref{ff}). By taking $\varepsilon= 10^{-2}$, the above values $(\tilde \rho_1,\tilde s_1)$ provide the following initial conditions (\ref{inicon})
{\small
\begin{equation}\label{inicondsqrt5}
\begin{split}
&(x(\varepsilon), y(\varepsilon),p_x(\varepsilon), p_y(\varepsilon))\ = \ (\,0.0182574,\, 2.00645,\, 0,\, 0\,)\ ,
\end{split}
\end{equation}
}
for the periodic orbit displayed in Fig. \ref{orbita2}. In this case $\omega_x/\omega_y=1/\sqrt{a^2-1}=1/2$, and the periodic orbit exhibits 1 oscillation in the  $x$ direction for every 2 oscillations in the $y$ direction. In this special case, two periodic orbits bifurcate from the equilibrium point $(0,\sqrt{a^2-1},0,0)$. The initial conditions of the second orbit (using $(\tilde \rho_2,\tilde s_2)$ ) are given by
{\small
\begin{equation*}
\begin{split}
&(x(\varepsilon), y(\varepsilon),p_x(\varepsilon), p_y(\varepsilon))\ = \ (\,0.0182574,\, 1.99355,\, 0,\, 0\,)\ .
\end{split}
\end{equation*}
}
The periodic orbits obtained from averaging \eqref{inicondsqrt5} are invariant under the symmetry $\mathcal{S}_1$. Of course, with the symmetry  $\mathcal{S}_2$ we obtain two additional periodic orbits from averaging theory,  bringing the total to four.

\vspace{-1.0cm}

\subsection{Case $a=\sqrt{29}/2$} 

For $h=1$, we obtain two zeros of the function (\ref{f}) such that $\dfrac{\partial(f_1,\,f_2)}{\partial(\rho,\, s)}\vert_{\rho=\tilde \rho,s=\tilde s}\,\neq \,0$, i.e.
\begin{equation*}
(\tilde \rho_1,\tilde s_1 )\ \approx \  (0.625998\,,\,-5.98141)\ ,
\end{equation*}
and
\begin{equation*}
(\tilde \rho_2,\tilde s_2 )\ \approx \ (0.625998\,,\,-3.58885)\ .
\end{equation*}

\begin{figure}[t]
	\centering
	\subfigure[$x$ \textit{vs} $p_x$]{\includegraphics[width=0.4\textwidth]{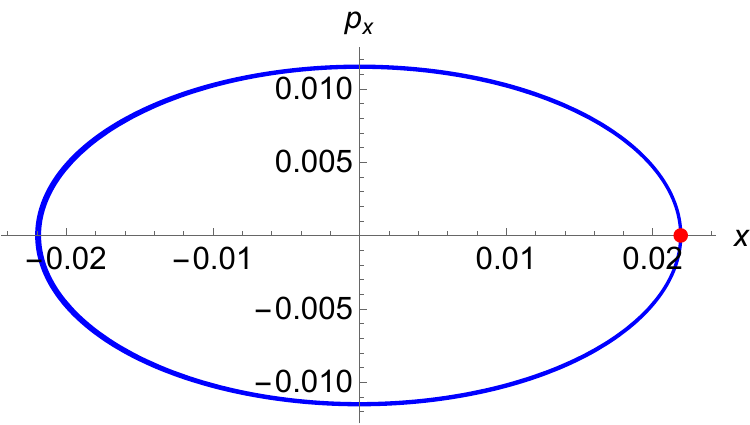}}
 	\subfigure[$x$ \textit{vs} $y$]{\includegraphics[width=0.42\textwidth]{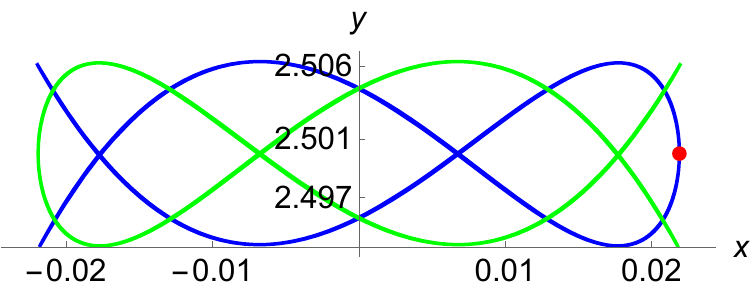}}
	\hfill
	\caption{Case $a=\frac{\sqrt{29}}{2}$. Periodic orbit (blue line) and its initial condition (\ref{inicondsqrt29})  marked by a red dot, obtained by averaging theory
 (a) on the plane $x-y$ (b) on the plane $x-p_x$. A simulation time of $20\,T$ was used in the computations. The two periodic orbits bifurcate from the equilibrium point $(0,\frac{5}{2},0,0)$. In (b) the second periodic orbit (in green) is obtained from the symmetry $(x,\,p_x) \rightarrow(-x,\, -p_x)$.}
	\label{orbita3}
\end{figure}

By taking $\varepsilon= 10^{-2}$ the above values $(\tilde \rho_1,\tilde s_1 )$ provide the following initial conditions (\ref{inicon}):
{\small
\begin{equation}\label{inicondsqrt29}
\begin{split}
&(\,x(\varepsilon), y(\varepsilon),p_x(\varepsilon), p_y(\varepsilon)\,)\ = 
\ (0.021911,\ 2.5,\ 0,\ 0.00822)\ ,
\end{split}
\end{equation}
}
which correspond to the periodic orbit displayed in Fig. \ref{orbita3}. In this case $\omega_x/\omega_y=1/\sqrt{a^2-1}=2/5$, and the periodic orbit has 2 oscillations in the  $x$ direction by 5 oscillations in the $y$ direction. From another side, the values $(\tilde \rho_2,\tilde s_2 )$ provide the initial conditions (\ref{inicon}):
{\small
\begin{equation}\label{inicondsqrt29}
\begin{split}
&(\,x(\varepsilon), y(\varepsilon),p_x(\varepsilon), p_y(\varepsilon)\,)\ = 
\ (0.021911,\ 2.5,\ 0,\ -0.00822)\ .
\end{split}
\end{equation} 
}
\bigskip

\section{On the non-integrability}\label{s6}

We consider the autonomous differential system
\begin{equation}\label{ea}
\dot x =f(x),
\end{equation}
where $f \colon U \to \R^n$ is $C^2$, $U$ is an open subset of $\R^n$ and the dot denotes the derivative with respect to the time~$t$. Let $x(t,x_0)$ be a periodic solution of the differential system \eqref{ea} of period $T$ such that $x(0,x_0)=x_0$.

The {\it variational equation} associated to the $T$-periodic solution
$x(t,x_0)$ is
\begin{equation}\label{eq:estrella}
\dot M= \bigg(\frac{\partial f(x)}{\partial x}\Big|_{x =x(t,x_0)} \bigg) M,
\end{equation}
where $M$ is an $n \times n$ matrix. Of course $\partial
f(x)/\partial x$ denotes the Jacobian matrix of $f$ with respect
to~$x$. 

The \emph{monodromy matrix} associated to the $T$-periodic solution $x(t,x_0)$ is the solution $M(T,x_0)$ of \eqref{eq:estrella} satisfying that $M(0,x_0)$ is the identity matrix. The eigenvalues of the monodromy matrix associated to the periodic solution $x(t,x_0)$ are called the \emph{multipliers} of the periodic orbit.

We recall an important theorem due to Poincaré \cite{Poincare}, (see also \cite{Koslov} p. 36 and \cite{LV}), on the integrability of a Hamiltonian system with two degrees of freedom.

\noindent{\bf Poincaré Theorem}. {\it If a Hamiltonian system with two degrees of freedom and Hamiltonian $H$ is Liouville--Arnold integrable, and $C$ is a second first integral such that the differentials of $H$ and $C$ are linearly independent at each point of a periodic orbit of the system, then
all the multipliers of this periodic orbit are equal to $1$.}

This theorem provides a tool for studying the non-integrability in the Liouville–Arnold sense, independently of the differentiability class of the second first integral $C$. The primary challenge in applying this theorem is finding periodic orbits with multipliers different from 1. The periodic orbits obtained in Theorem \ref{t1} and Corollary \ref{c1} are particularly useful for applying Poincaré's Theorem. In an autonomous Hamiltonian system, a periodic solution always has two multipliers equal to 1. One multiplier is 1 because the system is autonomous, and the other is 1 due to the existence of a first integral given by the Hamiltonian.

\begin{proof}[{\bf Proof of Theorem \ref{t2}}]
Under the hypotheses of Theorem \ref{t1} and Corollary \ref{c1}, the Jacobian determinant \eqref{deter} of the calculated periodic orbits depends on the value of the energy level $h$. The fundamental matrix associated with some of these periodic orbits is the product of the four multipliers of the orbit, with two of them always equal to 1. By appropriately selecting the values of $h$, the remaining two multipliers can be made different from 1. Thus, the Hamiltonian system is, in general, not Liouville–Arnold integrable because some multipliers are distinct from 1.

If all the determinants of the fundamental matrices associated to the periodic orbits of Corollary \ref{c1} are $1$, the Hamiltonian system can be  Liouville-Arnold integrable.
\end{proof}

It is important to emphasize the relation and differences of the present consideration with the following well-established theorems.

\begin{theorem}[Moser's Theorem]
If the function $f$ is of class $C^1$ and a first integral $H$ of system
\begin{equation}\label{1}
\dot x= f(x), \qquad x=(x_1,\ldots,x_m)
\end{equation}
is of class $C^2$ in a neighborhood of $x=0$, being $x=0$ an equilibrium point of this system, and asssuming that the gradient $H_x$  satisfies $H_x(0)=0$, and  that the Hessian matrix $H_{xx}(0)$ is positive definite, then for any sufficiently small $\varepsilon$ the energy surface $H(x)= H(0)+\varepsilon^2$ contains at least one periodic solution of system~\eqref{1} whose period is close to one of the linear differential system
\begin{equation*}
\dot x= Ax, \qquad A= f_x (0),
\end{equation*}
where $f_x(0)$ denotes the Jacobian matrix of the function $f$ evaluated at $x=0$.
\end{theorem}

\begin{theorem}[Weinstein's Theorem]
Consider the Hamiltonian system
\begin{equation}\label{2}
\dot x_k= H_{x_{n+k}}, \quad \dot x_{n+k}= -H_{x_k},\qquad k=1,\ldots,n,
\end{equation}
where $H_{x_l}$ denotes the partial derivative of the Hamiltonian $H(x_1,\ldots,x_{2n})$ with respect to the variable $x_l$. If the Hamiltonian $H$ is of class $C^2$ near $x=0$, $H_x(0)=0$, and the Hessian $H_{xx}(0)$ is positive definite, then for any sufficiently small $\varepsilon$ the energy surface $H(x)= H(0)+\varepsilon^2$ contains at least $n$ periodic solutions of system~\eqref{2} whose periods are close to those of the linear differential system with matrix the Jacobian matrix of system \eqref{2} at $x=0$.
\end{theorem}

Moser and Weinstein Theorems need that the Hessian $H_{xx}(0)$ be positive definite, and in general this is not the case for the Hamiltonian of the  two-center problem with harmonic-like interactions. So these theorems cannot be applied. On the other hand, in case that they could be applied they do not provide the approximate analytical expression of the periodic orbit given in Theorem 1 provided by the averaging theory.
\vspace{-15pt}

\section{Conclusions}

We have used the averaging theory for studying the periodic orbits of the Hamiltonian system modeling the two-center problem with harmonic-like interactions in some of their fixed Hamiltonian levels, see Theorem \ref{t1} and Corollary \ref{c1}. This tool can be applied to Hamiltonian systems with an arbitrary degrees of freedom.

Using a result due to Poincaré we have analyzed the non--integrability in the sense of Liouville--Arnold of the mentioned Hamiltonian systems, see Theorem \ref{t2}. Again this tool can be applied to Hamiltonian systems with an arbitrary number of degrees of freedom.

We must remark that these two tools are possible to apply if we have analytic information on the periodic orbits of the Hamiltonian system. Here this is the case thanks to the averaging theory for computing periodic orbits.

\subsection*{Credit authorship contribution statement}

The authors A M Escobar Ruiz, L Jiménez-Lara and Jaume Llibre have contributed equally to this paper (conceptualization, formal analysis, writing-original draft, validation). The author M A Zurita contributed with numerical computations of Poincaré sections and Lyapunov exponents.
\subsection*{Declaration of competing interest}

The authors declare that they have no known competing financial interests or personal relationships that could have appeared to influence the work reported in this paper.

\subsection*{Data availability}

No applicable.

\bigskip

\section*{Acknowledgments}

A.M. Escobar Ruiz would like to thank the support from Consejo Nacional de Humanidades, Ciencias y Tecnologías (CONAHCyT) of Mexico under Grant CF-2023-I-1496 and from UAM research grant DAI 2024-CPIR-0.

The third author is partially supported by the Agencia Estatal de Investigaci\'on of Spain grant PID2022-136613NB-100, AGAUR (Generalitat de Catalunya) grant 2021SGR00113, and by the Reial Acad\`emia de Ci\`encies i Arts de Barcelona.


\begin{thebibliography}{0}%
\makeatletter
\providecommand \@ifxundefined [1]{%
 \@ifx{#1\undefined}
}%
\providecommand \@ifnum [1]{%
 \ifnum #1\expandafter \@firstoftwo
 \else \expandafter \@secondoftwo
 \fi
}%
\providecommand \@ifx [1]{%
 \ifx #1\expandafter \@firstoftwo
 \else \expandafter \@secondoftwo
 \fi
}%
\providecommand \natexlab [1]{#1}%
\providecommand \enquote  [1]{``#1''}%
\providecommand \bibnamefont  [1]{#1}%
\providecommand \bibfnamefont [1]{#1}%
\providecommand \citenamefont [1]{#1}%
\providecommand \href@noop [0]{\@secondoftwo}%
\providecommand \href [0]{\begingroup \@sanitize@url \@href}%
\providecommand \@href[1]{\@@startlink{#1}\@@href}%
\providecommand \@@href[1]{\endgroup#1\@@endlink}%
\providecommand \@sanitize@url [0]{\catcode `\\12\catcode `\$12\catcode `\&12\catcode `\#12\catcode `\^12\catcode `\_12\catcode `\%12\relax}%
\providecommand \@@startlink[1]{}%
\providecommand \@@endlink[0]{}%
\providecommand \url  [0]{\begingroup\@sanitize@url \@url }%
\providecommand \@url [1]{\endgroup\@href {#1}{\urlprefix }}%
\providecommand \urlprefix  [0]{URL }%
\providecommand \Eprint [0]{\href }%
\providecommand \doibase [0]{http://dx.doi.org/}%
\providecommand \selectlanguage [0]{\@gobble}%
\providecommand \bibinfo  [0]{\@secondoftwo}%
\providecommand \bibfield  [0]{\@secondoftwo}%
\providecommand \translation [1]{[#1]}%
\providecommand \BibitemOpen [0]{}%
\providecommand \bibitemStop [0]{}%
\providecommand \bibitemNoStop [0]{.\EOS\space}%
\providecommand \EOS [0]{\spacefactor3000\relax}%
\providecommand \BibitemShut  [1]{\csname bibitem#1\endcsname}%
\let\auto@bib@innerbib\@empty
\end{thebibliography}%


\begin{thebibliography}{99}
	
\bibitem{Landau} L. Landau and E. Lifshitz,
\textit{Mechanics}, 3rd ed., Pergamon Press, Vol. 1,  1976.
	
\bibitem{GPS2002} H. Goldstein, C. Poole and J. Safko,
\textit{Classical Mechanics}, third edition. Addison-Wesley, 2002.
	
\bibitem{Arnold89} V. Arnold,
\textit{Mathematical Methods of Classical Mechanics}, second edition, Springer, 1989.
	
\bibitem{AKN2006} V. Arnold, V. Kozlov and A. Neishtadt,
\textit{Mathematical Aspects of Classical and Celestial Mechanics}, third edition, Springer, 2006.
	
\bibitem{Conto} Contopoulos G., Order and Chaos in Dynamical Astronomy, Springer Verlag, 2002.
	
\bibitem{Henon} M. Hénon and C. Heiles,
\textit{The applicability of the third integral of motion: Some numerical experiments}, Astron, J. 69 (1964), 73--79.

\bibitem{MacKay} MacKay, R.S. and Meiss, J.D., Hamiltonian Dynamical Systems, Taylor \& Francis, 1987.
	
\bibitem{Fatou} P. Fatou,
\textit{Sur le mouvement d'un syst{\`e}me soumis {\`a} des forces {\`a} courte p{\'e}riode}, Bull. Soc. Math. France 56 (1928), 98--139.
	
\bibitem{Bogoliubov1934} N.N. Bogoliubov and N. Krylov,
\textit{The application of methods of nonlinear mechanics in the theory of stationary oscillations}, Publ. 8 of the Ukrainian Acad. Sci. Kiev, 1934.
	
\bibitem{Bogoliubov1945} N.N. Bogoliubov,
\textit{Mathematical On some statistical methods in mathematical physics}, Izv. vo Akad. Nauk Ukr. SSR, Kiev, 1945.
	
\bibitem{Sanders} J.A. Sanders, F. Verhulst and J. Murdock,
\textit{Averaging Methods in Nonlinear Dynamical Systems}, Applied Mathematical Sciences, Second Edition , Springer New York, New York, 2007.
	
\bibitem{Verhulst} F. Verhulst,
\textit{Nonlinear Differential Equations and Dynamical Systems}, Universitext, Springer, 1991.
	
\bibitem{Adriana} A. Buică, J.P. Françoise  and J. Llibre,
\textit{Periodic solutions of nonlinear periodic differential systems with a small parameter}, Commun. Pure Appl. Anal. 6(1) (2007), 103--111.
	
\bibitem{Llibre_2011} J. Llibre and L. Jiménez-Lara,
\textit{Periodic orbits and non-integrability of Hénon–Heiles systems}, J. Phys. A: Math. Theor. 44 (2011), 205103-
	
\bibitem{Katz2019} O. Saporta Katz and E. Efrati,
\textit{Self-driven fractional rotational diffusion of the harmonic three-mass system}, Phys. Rev. Lett. 122 (2019), 024102.
	
\bibitem{3bodyexperimental} A.M. Escobar-Ruiz, M.A. Quiroz-Juarez and J.L. Del Rio-Correa,
\textit{Classical harmonic three-body system: an experimental electronic realization}, Sci Rep 12 (2022), 13346.
	
\bibitem{Katz2020}  O. Saporta Katz and E. Efrati,
\textit{Regular regimes of the harmonic three-mass system}, Phys. Rev. E 101  (2020), 032211.
	
\bibitem{Turbiner2020} A.V. Turbiner, W. Miller and  M.A. Escobar-Ruiz,
\textit{Three-body closed chain of interactive (an)harmonic oscillators and the algebra}, J. Phys. A 53 (2020), 055302.
	
\bibitem{Olivares-Pilon_2023} H. Olivares-Pilón, A.M. Escobar-Ruiz and F. Montoya Molina,
\textit{Three-body harmonic molecule}, J. Phys. B: At. Mol. Opt. Phys. 56 (2023), 075002.

\bibitem{BL} A. Buica and J. Llibre,
Averaging methods for finding periodic orbits via Brouwer degree, Bull. Sci. Math. 128 (2004), 7--22.

\bibitem{Ll} N.G. Lloyd,
\textit{Degree Theory}, Cambridge University Press, Cambridge, 1978.

\bibitem{Koslov}  V.V. Kozlov,
{\it Integrability and non-integrability in Hamiltonian mechanics}, Russian Math. Surveys 38 (1983), 1--76.

\bibitem{Poincare} H. Poincar\'e,
{\it Les m\'ethodes nouvelles de la m\'ecanique c\'eleste}, Vol. I,	Gauthier-Villars, Paris, 1899.

\bibitem{LV} J. Llibre and C. Valls,
{\it On the $C^1$ non-integrability of differential systems via periodic orbits}, European J. Appl. Math. {\bf 22} (2011), 381--391.
	
\end{thebibliography}
\end{document}